\documentclass[letterpaper,prc,twocolumn,showpacs,floatfix,nofootinbib,preprintnumbers,superscriptaddress,amsmath,amssymb]{revtex4-1}

\usepackage{amsmath}           
\usepackage{amsfonts}
\usepackage{graphicx}          
\usepackage{dcolumn}           
\usepackage{bm}

\usepackage{color}
\newcommand{\xb}{\boldsymbol{x}}  

\newcommand{\xh}{\hat{\xb}}  

\newcommand{\algo}{\textsc{pound}er\textsc{s}}

\newcommand{\HFODD}{\textsc{hfodd}}
\newcommand{\HFBTHO}{\textsc{hfbtho}}

\newcommand{\UNEDFZERO}{\textsc{unedf0}}
\newcommand{\UNEDFNB}{\textsc{unedf}nb}
\newcommand{\exclude}[1]{}

{\catcode`\|=\active
  \gdef\Braket#1{\left<\mathcode`\|"8000\let|\bravert {#1}\right>}}
\newcommand{\bravert}{\egroup\,\vrule\,\bgroup}

\newcommand{\enm}{E}
\newcommand{\rhoc}{\rho_\text{c}}
\newcommand{\knm}{K}

\newcommand{\asym}{a_\text{sym}}

\newcommand{\lsym}{L}

\newcommand{\gras}[1]{\boldsymbol{#1}}

\newcommand{\ba}{\begin{array}}
\newcommand{\ea}{\end{array}}
\newcommand{\disregard}[1]{}

\newcommand{\UNEDFONE}{\textsc{unedf1}}
\newcommand{\UNEDF}{\textsc{unedf}}

\newcommand{\UNEDFTWO}{\textsc{unedf2}}

\begin{document}

\preprint{\fbox{\sc version of \today}}

\title{Nuclear energy density optimization: Shell structure}

\author{M.~Kortelainen}
\affiliation{Department of Physics, University of Jyv\"askyl\"a, P.O. Box 35 (YFL), FI-40014 Finland}
\affiliation{Department of Physics and Astronomy, University of Tennessee,
Knoxville, TN 37996, USA}
\affiliation{Physics Division, Oak Ridge National Laboratory, Oak Ridge, TN
37831, USA}

\author{J.~McDonnell}
\affiliation{Department of Physics and Astronomy, University of Tennessee,
Knoxville, TN 37996, USA}
\affiliation{Physics Division, Oak Ridge National Laboratory, Oak Ridge, TN
37831, USA}
\affiliation{Physics Division, Lawrence Livermore National Laboratory,
Livermore, CA 94551, USA}

\author{W.~Nazarewicz}
\affiliation{Department of Physics and Astronomy, University of Tennessee,
Knoxville, TN 37996, USA}
\affiliation{Physics Division, Oak Ridge National Laboratory, Oak Ridge, TN
37831, USA}
\affiliation{Institute of Theoretical Physics, Warsaw University, ul. Ho\.{z}a 69, PL-00681, Warsaw, Poland}

\author{E.~Olsen}
\affiliation{Department of Physics and Astronomy, University of Tennessee,
Knoxville, TN 37996, USA}

\author{P.-G.~Reinhard}
\affiliation{Institut f\"ur Theoretische Physik, Universit\"at Erlangen, D-91054 Erlangen, Germany}

\author{J.~Sarich}
\affiliation{Mathematics and Computer Science Division, Argonne National Laboratory, Argonne, IL 60439, USA}

\author{N.~Schunck}
\affiliation{Physics Division, Lawrence Livermore National Laboratory,
Livermore, CA 94551, USA}
\affiliation{Department of Physics and Astronomy, University of Tennessee,
Knoxville, TN 37996, USA}
\affiliation{Physics Division, Oak Ridge National Laboratory, Oak Ridge, TN
37831, USA}

\author{S.~M.~Wild}
\affiliation{Mathematics and Computer Science Division, Argonne National Laboratory, Argonne, IL 60439, USA}

\author{D.~Davesne}
\affiliation{Universit\'{e} de Lyon, F-69622 Lyon, France; Universit\'{e} de Lyon 1, Villeurbanne; CNRS/IN2P3, Institut de Physique Nucl\'{e}aire de Lyon}

\author{J.~Erler}
\affiliation{Division of Biophysics of Macromolecules, German Cancer Research Center (DKFZ), Im Neuenheimer Feld 580, D-69120 Heidelberg, Germany}

\author{A.~Pastore}
\affiliation{Institut d'Astronomie et d'Astrophysique, Universit\'{e} Libre de Bruxelles - CP226, 1050 Brussels, Belgium}

\date{\today}

\begin{abstract}
\begin{description}
\item[Background]
Nuclear density functional theory is the only microscopical theory that can be
applied throughout the entire nuclear landscape. Its key ingredient is the
energy density functional.
\item[Purpose]
In this work,  we propose a new parameterization {\UNEDFTWO} of the
Skyrme energy density functional.
\item[Methods]
The functional optimization is carried out using  the {\algo} optimization
algorithm within the framework of the Skyrme Hartree-Fock-Bogoliubov theory.
Compared to the previous parameterization {\UNEDFONE}, restrictions on the
tensor term of the energy density have been lifted, yielding a very general
form of the energy density functional up to second order in derivatives
of the one-body density matrix. In order to impose constraints on all the
parameters of the functional, selected data on single-particle splittings in
spherical doubly-magic nuclei have been included into the experimental dataset.
\item[Results]
The agreement with both bulk and spectroscopic nuclear properties achieved by
the resulting {\UNEDFTWO} parameterization is comparable with {\UNEDFONE}.
While there is a small improvement on single-particle spectra and binding
energies of closed shell nuclei, the reproduction of fission barriers and
fission isomer excitation energies has degraded. As compared to previous
{\UNEDF} parameterizations, the parameter confidence interval for {\UNEDFTWO}
is narrower. In particular, our results overlap well with those obtained in
previous systematic studies of the spin-orbit and tensor terms.
\item[Conclusions]
{\UNEDFTWO} can be viewed as an all-around Skyrme EDF that performs reasonably
well for both global nuclear properties and shell structure. However, after
adding new data aiming to better constrain the nuclear functional, its quality
has improved only marginally. These results suggest that the standard Skyrme energy
density has reached its limits and significant changes to the form
of the functional are needed.
\end{description}
\end{abstract}


\maketitle


\section{Introduction}
\label{sec:introduction}

An important goal in research in low-energy nuclear physics is to develop
an universal nuclear energy density functional (EDF) that can be used to
explain and predict static and dynamic properties of atomic nuclei within the
framework of nuclear density functional theory (DFT). Building such a
functional has been one of the primary drivers behind the formation of the
former UNEDF SciDAC-2 collaboration \cite{(Ber07),*(Fur11),*(Nam12),(Bog13)};
its current successor, the NUCLEI SciDAC-3 collaboration \cite{NUCLEI}; and the
FIDIPRO collaboration \cite{FIDIPRO}.

Most of the nuclear EDFs used in self-consistent mean-field calculations have
been derived from phenomenological effective interactions, or pseudopotentials
\cite{(Vau72),(Neg72),(Ben03)}. A recent promising development is to use
results from effective field theory combined with density matrix expansion
techniques to construct a realistic EDF based on chiral interactions
\cite{(Lal04),(Geb10),(Car10),(Sto10),(Bog11)}. In a parallel effort,
methodologies have been developed to validate nuclear EDFs by optimizing their
low-energy coupling constants to experimental data on finite nuclei and
pseudo-data on nuclear matter and other relevant systems
\cite{(Bog13),(Klu09),(Kor10),(Kor12),(Erl13)}. One of the main challenges of
EDF optimization is to find the most relevant fit observables that can tightly
constrain the parameter space of the model. This requires a careful analysis of
the contribution of each term of the functional to low-energy nuclear
properties. In finite nuclei, the most common observables used in EDF fits are
binding energies and their differences, charge radii, surface thickness, and
energies of giant resonances \cite{(Ben03),(Sto07)}. Given a mathematical form
of the EDF, its predictive power ultimately depends on the choice of the data
used in the optimization. In particular, applications of DFT to nuclear
spectroscopy are very sensitive to the details of shell structure; this
requires a careful choice of fit observables.

Shell structure is the fundamental property of the atomic nucleus
\cite{(Boh75)}. In an independent-particle  picture, shell structure can be
associated with the single-particle (s.p.) spectra of the  mean-field potential
\cite{(Rin00),(Nil95)}. Reproducing the correct ordering and distribution of
s.p.\ levels is, therefore, an essential requirement for nuclear structure
theories, but it has to be approached with caution
\cite{(Pan97),(Rin00),(Dug12)} since s.p.\ motion is significantly modified by
correlations \cite{(Col10),(Dug12),(Tar13)}. In the context of nuclear DFT, many commonly used  EDFs have been
optimized by explicitly using some experimental input pertaining to the s.p.\ 
level structure in doubly-magic nuclei \cite{(Ben03)}.

The s.p.\ shell structure is very sensitive to the details of the effective
interaction or the energy density and is the result of a subtle interplay
between the gradient terms and effective mass, spin-orbit, and tensor terms
\cite{(Sat08),(Kor08)}. Suggestions to study tensor interactions within the
self-consistent mean-field approach were made already in the seventies
\cite{(Sta77)} but the limited experimental data available did not provide
sufficient sensitivity to adjust the related coupling constants. In recent
years, the role of tensor coupling constants, in Skyrme EDFs in particular, has
been thoroughly investigated
\cite{(Bro06),(Dob06c),*(Dob07b),(Les07),(Zou08),(Tar08),(Kor08),(Sat08),(Zal08),(Zal09),(Ben09b),(Mor10),(Wan11),(Gra13)}.
An important conclusion from several of those papers is that the inclusion of
tensor terms should not be done perturbatively but should instead involve the
complete EDF reoptimization at the deformed Hartree-Fock-Bogoliubov (HFB)
level. This implies that constraints on the tensor terms must be included in
the pool of fit observables.

In our previous works on Skyrme EDF optimization \cite{(Kor10),(Kor12)}, tensor
terms were disregarded because our dataset did not contain any information
specifically constraining shell structure. This limitation is lifted in this article,
which should be viewed as the continuation of our work on energy density
parameter optimization. In Ref.~\cite{(Kor10)}, we presented the main strategy
underlying our optimization protocol and developed the {\UNEDFZERO} EDF
parameterization by using experimental input on a selected set of nuclear
masses, charge radii, and odd-even mass differences. In the same paper, we
performed one of the first sensitivity analyses of Skyrme parameterizations to
obtain the correlations and standard deviations for the parameters. In
Ref.~\cite{(Kor12)}, we modified the form of the functional by removing the
center of mass correction, which allows for straightforward time-dependent
Hartree-Fock (HF) and HFB applications. To constrain deformation properties, we
also extended our dataset to include information  on fission isomer excitation
energies. In this way, nuclear deformation properties produced by the resulting {\UNEDFONE}
functional have greatly improved, in particular in the context of nuclear
fission.

The goal of this study is to include the tensor coupling constants in the set
of optimized parameters. The resulting energy density is a general
functional of the one-body density matrix up to second order in derivatives. Because
shell structure is very sensitive to the tensor terms of the functional, we extend our
experimental dataset by adding a set of s.p.\ energy splittings in doubly-magic
nuclei. The optimization of the functional within this extended  dataset yields
the {\UNEDFTWO} parameterization of the Skyrme energy density. In the spirit of
our previous work, we carry out a full sensitivity analysis of {\UNEDFTWO},
which is essential to assessing the predictive power of the theory
\cite{(Rei10),(Fat11),(Erl12),(Gao13),(Kor13),(Rei13),(Rei13a),(Naz13)}. Since
the previous work of Ref.~\cite{(Kor08)} based on the linear regression
methodology demonstrated that the current standard form of the Skyrme EDF
cannot ensure a spectroscopic-quality description of s.p.\ energies, {\UNEDFTWO}
is certainly not the universal nuclear EDF. It can be viewed, however, as the
best all-around Skyrme EDF that performs reasonably well for both global
nuclear properties and shell structure. For this reason, we consider
{\UNEDFTWO} as the end of the standard Skyrme EDF journey.

This paper is organized as follows. In Sec.~\ref{sec:theory}, we briefly review
the theoretical framework and the notations. Section~\ref{sec:optimization}
describes the optimization method employed, experimental data used in the fit,
and presents the {\UNEDFTWO} parameterization together with its sensitivity
analysis. Global nuclear properties computed with {\UNEDFTWO} are reviewed in
Sec.~\ref{sec:results}. Finally, Sec.~\ref{sec:conclusions} contains
conclusions and perspectives for future work.


\section{Theoretical Framework}
\label{sec:theory}

In the nuclear DFT, the total energy $E$ is a functional of the one-body
density matrix $\rho$ and pairing density $\tilde\rho$ and
can be cast into the generic form
\begin{eqnarray}\label{eq:EDFT}
E[\rho,\tilde{\rho}] & = & \int d^{3}\gras{r} ~
\left[
{\mathcal E}_{\rm Kin}(\gras{r}) +\chi_0(\gras{r}) +\chi_1(\gras{r})
\right. \nonumber \\ & & \left.
+ \tilde{\chi}(\gras{r})+{\mathcal E}_{\rm Coul}(\gras{r})
\right],
\end{eqnarray}
where  ${\mathcal E}_{\rm Kin}(\gras{r})$ is the kinetic energy;
$\chi_t(\gras{r})$ is the isoscalar ($t=0$) and isovector ($t=1$) particle-hole
Skyrme energy density; $\tilde{\chi}(\gras{r})$  is the pairing energy density;
and ${\mathcal E}_{\rm Coul}(\gras{r})$ is the Coulomb term.

The particle-hole part of the Skyrme energy density reads
\begin{eqnarray}
\chi_t(\gras{r}) & = &
C_t^{\rho\rho} \rho_t^2  +
C_t^{\rho\tau} \rho_t\tau_t +
C_t^{JJ} \sum_{\mu\nu} J_{\mu\nu,t}J_{\mu\nu,t}
\nonumber \\  &    &
+ C_t^{\rho\Delta\rho} \rho_t\Delta\rho_t +
C_t^{\rho \nabla J} \rho_t\bm{\nabla}\cdot\bm{J}_t,
\label{eq:UED}
\end{eqnarray}
where each term is multiplied by a coupling constant $C_{t}^{uu'}$ represented
by a real number. The coupling constant $C_t^{\rho\rho}$ is the only exception,
as it has the traditional density-dependence
\begin{equation}
C_t^{\rho\rho} =  C_{t0}^{\rho\rho}  + C_{t{\rm D}}^{\rho\rho}~ \rho_0^\gamma \, .
\label{eq:crramp}
\end{equation}
The definitions of the various densities $\rho$, $\tau$, and $J_{\mu\nu}$
(${\bm J}$ is the vector part of $J_{\mu\nu}$) can be found in
Ref.~\cite{(Eng75),(Dob96),(Ben03),(Per04),(Les07)}. The coupling constants
$C_t^{\rho\rho}$ and $C_t^{\rho\tau}$ are related to the volume part of the
energy density and can be expressed as a function of the parameters of infinite
nuclear matter \cite{(Kor10)}.

The term $\sum_{\mu\nu} J_{\mu\nu,t}J_{\mu\nu,t}$ in Eq.~(\ref{eq:UED})
represents a tensor energy density. The approximation that was made in the
{\UNEDFZERO} and {\UNEDFONE}  optimizations was to set $C_t^{JJ} = 0 $ for
both $t=0$ and $t=1$. In the present work, this constraint has been removed,
and these two tensor coupling constants are taken as free parameters. Note that
we do not allow independent variations of the pseudoscalar, vector, and
peudotensor components of $J_{\mu\nu,t}$: each of these components is
multiplied by the same coupling constant, see \cite{(Dob96),(Per04)}.

The pairing term is derived from the mixed pairing force of
Ref.~\cite{(Dob02a)}, leading to the energy density
\begin{equation}
\tilde{\chi}(\gras{r}) =
\frac{1}{4}\sum_{q={\rm n,p}}
V^{q}_{0}
\left[ 1 - \frac{1}{2}\frac{\rho_{0}(\gras{r})}{\rho_{\rm c}} \right]
\tilde\rho^2(\gras{r}),
\label{eq:vpair}
\end{equation}
where  $V^{q}_{0}$ $(q=n,p)$  is the pairing strength. In this work as before,
we take $\rho_{\rm c}$=0.16\, fm$^{-3}$. We have used different pairing
strengths for neutrons and protons \cite{(Ber09a)}. Owing to the zero range of
our effective pairing force, we have used a pairing cut-off
$E_{\rm cut}=60$\,MeV to truncate the quasi-particle space. To prevent the
collapse of pairing correlations near closed shells, we have also used the
variant of the Lipkin-Nogami (LN) method as in  Ref.~\cite{(Sto03)}.

As with {\UNEDFONE}, we disregard the center-of-mass correction; for motivation
see Ref.~\cite{(Kor12)}. The Coulomb term  contains direct and exchange
contributions. The direct part is computed from the proton density distribution
assuming a point-proton charge, and the exchange part is treated at the Slater
approximation.


\section{Optimization of Energy Density and Sensitivity Analysis}
\label{sec:optimization}

In this section, we present our optimization protocol and analyze the features
of the resulting {\UNEDFTWO} parameterization. In
Sec.~\ref{subsec:experimental},
we review the experimental data used in the fit. In Sec.~\ref{subsec:blocking},
we describe how information on shell structure was incorporated. The
optimization procedure itself is summarized in
Sec.~\ref{subsec:optimization}. Section~\ref{subsec:UNEDF2} compares
{\UNEDFTWO} to previous {\UNEDF} parameterizations, and
Sec.~\ref{subsec:sensitivity} contains the results of the sensitivity analysis.


\subsection{Experimental Dataset}
\label{subsec:experimental}

In order to determine the tensor coupling constants $C_t^{JJ}$, one must
properly select the experimental fit observables to effectively constrain their
values. To this end, we have extended the previous dataset used in the
{\UNEDFONE} optimization by including an additional nine single-particle level
splittings, five new data points for odd-even staggering (OES), and one
additional binding energy.

\begin{table}[!htb]
\caption{Empirical single-particle level splittings \cite{(Sch07),(Oros96)} (in
MeV) used in the {\UNEDFTWO} optimization. The labels $n$ and $p$ refer to the
neutron and proton levels, respectively.}
\begin{ruledtabular}
\begin{tabular}{cccc}
Nucleus & $n/p$ & Level & Energy \\
\hline  \\[-8pt]
$^{40}$Ca  & $n$ & $f_{5/2}-f_{7/2}$   & 6.80 \\
$^{40}$Ca  & $n$ & $f_{7/2}-d_{3/2}$   & 7.28 \\
$^{40}$Ca  & $p$ & $f_{7/2}-d_{3/2}$   & 7.24 \\
$^{48}$Ca  & $n$ & $f_{5/2}-f_{7/2}$   & 8.80 \\
$^{48}$Ca  & $p$ & $f_{5/2}-f_{7/2}$   & 4.92 \\
$^{132}$Sn & $n$ & $h_{9/2}-h_{11/2}$  & 6.68 \\
$^{132}$Sn & $p$ & $g_{7/2}-g_{9/2}$   & 6.03 \\
$^{208}$Pb & $n$ & $i_{11/2}-i_{13/2}$ & 6.08 \\
$^{208}$Pb & $p$ & $h_{9/2}-h_{11/2}$  & 5.56
\end{tabular}
\end{ruledtabular}
\label{table:newspdata}
\end{table}

We show in Table \ref{table:newspdata} the empirical values of single-particle
splittings in several doubly-magic nuclei. All values are taken from the empirical s.p. energies listed in Ref.~\cite{(Sch07)} except for the proton  $f_{5/2}-f_{7/2}$ splitting in
$^{48}$Ca, which is taken from Ref.~\cite{(Oros96)}. The rationale to use
s.p.\ splittings instead of the absolute energy of levels is to remove some of
the systematic errors induced by the use of a truncated harmonic oscillator (HO) basis. We set the weight of the single-particle data points in the $\chi^{2}$
function to $w=1.2$\,MeV. This choice was motivated based on the singular value
decomposition (SVD) analysis performed in Ref.~\cite{(Kor08)}, which showed
that Skyrme EDFs can reproduce empirical s.p.\ levels at this precision level.
We recall that the weight can be viewed as a coarse estimate of the theoretical error on a given observable.

The calculation of s.p.\ splittings in $^{132}$Sn requires the ground-state
energy of $^{132}$Sn, see Sec.~\ref{subsec:blocking}. For this reason, we added
the binding energy of this nucleus to the dataset. As in Ref.~\cite{(Kor10)},
experimental information has been taken from the 2003 mass evaluation, and the
nuclear binding energy was obtained after taking into account the electronic
correction, yielding the value $B(^{132}{\rm Sn})=-1102.686066$\,MeV. For this
additional datum, we took the same weight $w=2$\,MeV as for other binding
energies.

\begin{table}[!htb]
\caption{New data for  $\tilde{\Delta}^{(3)}_{q}$  (in MeV) used in
{\UNEDFTWO} optimization.}
\begin{ruledtabular}
\begin{tabular}{rrrrrr}
\multicolumn{3}{c}{Neutrons} & \multicolumn{3}{c}{Protons} \\
$Z$ & $N$ & $\tilde{\Delta}^{(3)}_{\rm n}$ & $Z$ & $N$ & $\tilde{\Delta}^{(3)}_{\rm p}$ \\
\hline
90 & 142 & 0.681450 & 90 & 142 & 0.813287 \\
50 &  74 & 1.250400 & 76 &  90 & 1.169046 \\
50 &  70 & 1.316825 & \\
\end{tabular}
\end{ruledtabular}
\label{table:newdeldata}
\end{table}

In addition to the single-particle splittings and the binding energy of
$^{132}$Sn, we have added five new OES data points, which are listed in Table
\ref{table:newdeldata}. This was motivated by the observation that pairing
properties of actinide nuclei and neutron-rich tin isotopes are poorly
reproduced by {\UNEDFONE}, suggesting that the weight of pairing-related data
in the objective function should be increased. We recall that the experimental
OES that we use is defined as the average of two odd-even (protons) and
even-odd (neutrons) $\Delta^{(3)}_{q}$ values, that is,
$\tilde\Delta^{(3)}_{n}(N)=\left[\Delta^{(3)}(N-1)+\Delta^{(3)}(N+1)\right]/2$.
In addition to these new experimental points, we have increased the weight of
all OES data points in the optimization from $w=0.050\,{\rm MeV}$ to
$w=0.100\,{\rm MeV}$.

To summarize, the {\UNEDFTWO} optimization dataset contains 47 deformed binding
energies, 29 spherical binding energies, 28  proton point radii, 13 OES values,
4 fission isomer excitation energies, and 9 single-particle level splittings.
The changes with respect to the  {\UNEDFONE} optimization procedure are as
follows:
\begin{itemize}
\item Tensor coupling constants $C_{0}^{JJ}$ and $C_{1}^{JJ}$ are
optimized;
\item The binding energy of $^{132}$Sn is added to dataset;
\item 9 new single-particle splittings are included in dataset;
\item 5 new OES data points are added to dataset;
\item The weight of all OES data points is increased to $0.1\,{\rm MeV}$.
\end{itemize}


\subsection{Computation of Single-Particle Levels}
\label{subsec:blocking}

Most of the optimizations of the Skyrme EDF that included information on s.p.
splittings were conducted at the spherical HF level (see, for
example, Refs.~\cite{(Bro98),(Cha98),(Klu09)}). Within this approximation, the
many-body wave-function reduces to a single Slater determinant, and the
theoretical single-particle levels are taken as the eigenvalues of the HF
Hamiltonian $\varepsilon_{j}$ following Koopmans' theorem \cite{(Koo34),(Kor08)}.
In the framework of the HFB theory with approximate particle number projection,
where the basic degrees of freedom are not particles but quasi-particles,
separation energies provide a more convenient quantity to relate to effective
single-particle energies.

In our optimization procedure, theoretical single-particle splittings were thus
obtained by applying the blocking HFB approach with the approximate
LN correction. We employ the equal-filling approximation to
blocking, since it yields an excellent estimate of the full symmetry-breaking
blocking results \cite{(Sch10)}. Since the shape polarization induced by the
blocking prescription spontaneously breaks the spherical symmetry of the
odd-$A$ nucleus, the energy degeneracy of a blocked spherical quasi-particle
orbital with angular momentum $j$ is lifted \cite{(Zal08),(Sch10),(Sat12)}. In
the equal filling approximation, however, time-reversal symmetry is conserved,
and states with the angular momentum projection +$\Omega$ and -$\Omega$ are
degenerate.

This fragmentation of any given $j$-shell in non-spherical blocking calculations poses a practical difficulty. Indeed, one should in principle compare the energy of all obtained blocking configurations with different $|\Omega|$ values, and pick the lowest one to compare with experiment.
The difficulty with such a strategy is that a configuration
with $|\Omega| < j$ can originate from a $j$-shell that is different from the
one under consideration. To avoid such a situation, we have chosen to block the
single state with the maximum projection $\Omega=+j$. The associated systematic
error does not exceed 100\,keV \cite{(Kor12)}.

Empirical s.p. energies are usually extracted from the centroids of (often
broad) strength functions of pick-up/stripping reactions. In our approach, we
{\em choose} to relate these empirical levels to one-particle separation
energies computed at the HFB+LN approximation. This choice allows us to remain
consistent throughout and calculate all observables at the same approximation
level (HFB+LN). In addition, we find that empirical energy splittings extracted
from the s.p. energies of the even-even nucleus or directly from the separation
energy of the odd nucleus in a given $J^{\pi}$ configuration differs typically
by at most a few hundreds keV, with one notable exception.
This should be compared with the $>$1.2 MeV accuracy of Skyrme
functionals for s.p. data, and suggests that the determination of the empirical
value should not have a large impact on the optimization.
The exception for the neutron $f_{5/2}$ state in $^{40}$Ca is due to the strong
fragmentation of the $f_{5/2}$ strength among multiple states, resulting in a rather
broad centroid \cite{(Uoz94)}. Similarly, the proton $f_{5/2}$ state in
$^{48}$Ca is fragmented among multiple states \cite{(Bur08)}.
In the Supplemental Material at \cite{SM}, we provide both sets of experimental s.p.
splittings for convenience.

Our procedure to generate  theoretical s.p.\ splittings follows that of
Ref.~\cite{(Rut98)} and can be summarized as follows. We begin by computing
a reference spectrum in the doubly magic nucleus of interest. Next, we use this
quasi-particle  spectrum to identify the blocking configuration with
$\Omega = +j$ and perform the blocking calculation in the system with $\pm 1$
particles. The effective particle and hole s.p.\ energies are respectively
defined as
\begin{subequations}
\begin{eqnarray}
E_{\rm s.p.}^{(\text{part.})} & = & E_{\rm bl}(A+1)-E(A), \label{eq:spEpart} \\
E_{\rm s.p.}^{(\text{hole})} & = & E(A)-E_{\rm bl}(A-1), \label{eq:spEhole}
\end{eqnarray}
\end{subequations}
where $A$ refers to the particle number of the reference (doubly-magic) nucleus
and $E_{\rm bl}$ is the energy of the blocked configuration in the
neighboring odd nucleus. The labels ``hole'' and ``particle'' refer to whether
the corresponding s.p.\ levels would be, respectively, fully occupied or empty
in the corresponding HF calculation of the doubly-magic nucleus. In the case where the s.p.\ 
levels involved in the s.p.\ splitting are both either above or below the Fermi
surface, the contribution of the even-even binding energy cancels out, and the
s.p.\ splitting reduces to the difference of total binding energies of blocked
configurations.


\subsection{Optimization}
\label{subsec:optimization}

All HFB calculations in the {\UNEDFTWO} optimization were performed with the
DFT solver {\HFBTHO} \cite{(Sto13)}. The code solves the HFB equations in an
axially symmetric deformed HO basis. In our initial work
on {\UNEDFZERO}, we used a spherical HO basis with 20 shells and assumed the
HFB solution to be reflection symmetric. Adding experimental data on fission
isomer excitation energies in the {\UNEDFONE} optimization required computing
the energy of super-deformed (SD) configurations. In order to mitigate
truncation errors, the SD states were calculated with a deformed, or stretched,
HO basis with the axial quadrupole deformation parameter $\beta=0.4$. In this work,
we have maintained the same setup as for {\UNEDFONE}: the spherical basis is
used for all ground-state configurations, and the stretched basis with
$\beta = 0.4$ is used for SD states. In all cases, the spherical frequency
$\omega_{0}$ of the HO basis is set at $\hbar\omega_{0} = 41/A^{1/3}$
\cite{(Sto13)}.

The objective function in our optimization is
\begin{equation}
\chi^{2}(\gras{x}) =
\frac{1}{n_{\rm d}-n_{\rm x}}
\sum_{i=1}^{D_{T}}
\sum_{j=1}^{n_{i}}
\left(\frac{s_{i,j}(\gras{x})-d_{i,j}}{w_{i}}\right)^{2} \, ,\label{eq:chi2}
\end{equation}
where $D_{T}$ is the number of different data types; for {\UNEDFTWO}
$D_{T} = 5$. The total number of data points is
$n_{d} = \sum_{i=1}^{D_{T}} n_{i}$ (here $n_{d} = 130$), and the number of
parameters to be fitted is $n_{x}$ (here $n_{x} = 14$). The calculated value of
the $j^{\mathrm{th}}$ observable of type $i$ is $s_{i,j}(\gras{x})$, while the
corresponding experimental value is denoted by $d_{i,j}$. Each data type has a
weight $w_{i}$. As in the cases of {\UNEDFZERO} and {\UNEDFONE}, the isovector
effective mass $1/M^{*}_{\rm v}$ is kept constant during the optimization,
since we find that it cannot be constrained reliably with the current data; we
retain the SLy4 value for $1/M^{*}_{\rm v}$ for historical reasons \cite{(Cha98)}. In the Supplemental Material  \cite{SM}, we provide the full list of experimental data
points that were used in all three optimizations: {\UNEDFZERO}, {\UNEDFONE},
and {\UNEDFTWO}.

As in our previous work, the parameters of the functional are not allowed to
attain unphysical values: we impose bounds on the range of variation of each
parameter. Bounds for the parameters common to both {\UNEDFONE} and {\UNEDFTWO}
were assumed to be the same. We did not set any bounds on the tensor coupling
constants $C_{t}^{JJ}$. During the optimization, two of the parameters,
$\enm/A$ and $\lsym$, ran to their boundary and were fixed to those values. They
were subsequently  excluded from the sensitivity analysis. The optimization was
carried out with the same {\algo} algorithm that was used for other {\UNEDF}
parameterizations, see Ref.~\cite{(Kor10)} for details.


\subsection{The UNEDF2 parameterization}
\label{subsec:UNEDF2}

The  optimized parameter set of the EDF {\UNEDFTWO} is listed in Table
\ref{table:unedf2} along with the standard deviation of each parameter and the
95\% confidence intervals. To facilitate legibility, we  display only the first
few significant digits of each parameter. In the Supplemental Material \cite{SM} we provide the parameter values of all three parameterizations up to machine
precision in two different representations: the hybrid nuclear
matter/coupling constants representation and the full coupling constant
representation.

\begin{table}[!htb]
\caption{Values $\hat{\xb}$ of the  Skyrme functional {\UNEDFTWO} parameters
$\xb$. Listed are final optimized parameter values, standard deviations, and
95\% confidence intervals. $ \rhoc $ is in fm$^{-3}$; $ \enm/A $, $ \knm $,
$ \asym $, and  $ \lsym $ are in MeV; $ 1/M_{s}^{*} $  is dimensionless;
$ C_{t}^{\rho\Delta\rho} $, $ C_{t}^{\rho\nabla J}$, and $ C_{t}^{JJ}$ in
MeV\,fm$^5$; and $V_0^n$ and $ V_0^p $ in  MeV\,fm$^3$.}
\label{table:unedf2}
\begin{ruledtabular}
\begin{tabular}{lcccc}
$ \xb $                       & $ \hat{\xb}^{\rm (fin.)} $ & $\sigma$ & 95\% CI \\
\hline
$ \rhoc $                  &    0.15631 & 0.00112 & [   0.154,   0.158] \\
$ \enm/A $                 &  -15.8     &      -  &          -          \\
$ \knm $                   &  239.930   & 10.119  & [ 223.196, 256.663] \\
$ \asym $                  &   29.131   &  0.321  & [  28.600,  29.662] \\
$ \lsym $                  &   40.0     &     -   &          -          \\
$ 1/M_{s}^{*} $            &    1.074   &  0.052  & [   0.988,   1.159] \\
$ C_{0}^{\rho\Delta\rho} $ &  -46.831   &  2.689  & [ -51.277, -42.385] \\
$ C_{1}^{\rho\Delta\rho} $ & -113.164   & 24.322  & [-153.383, -72.944] \\
$ V_0^n $                  & -208.889   &  8.353  & [-222.701,-195.077] \\
$ V_0^p $                  & -230.330   &  6.792  & [-241.561,-219.099] \\
$ C_{0}^{\rho\nabla J}$    &  -64.309   &  5.841  & [ -73.968, -54.649] \\
$ C_{1}^{\rho\nabla J}$    &  -38.650   & 15.479  & [ -64.246, -13.054] \\
$ C_{0}^{JJ}$              &  -54.433   & 16.481  & [ -81.687, -27.180] \\
$ C_{1}^{JJ}$              &  -65.903   & 17.798  & [ -95.334, -36.472] \\
\end{tabular}
\end{ruledtabular}
\end{table}

We first note that for {\UNEDFTWO}, the nuclear incompressibility parameter
$\knm$, while in the top range of acceptable values, is now constrained by the
data, whereas the slope of the symmetry energy $\lsym$ is not. As expected,
both the neutron and pairing strengths are also a little larger, a direct
consequence of adding more OES points into the dataset.

Table \ref{table:unedfs} lists all three {\UNEDF} parameterizations produced so
far, and compares them to the SLy4 parametrization, which was the starting
point for {\UNEDFZERO}. Interestingly, the {\UNEDFTWO} and {\UNEDFONE}
parameterizations are quite similar overall. This result is a little
surprising: one may have expected that relaxing the constraints on the tensor
coupling constants would lead to a significant rearrangement of all other
coupling constants, in particular the spin-orbit coupling constants. Indeed, it
was shown in Ref.~\cite{(Les07)} that there is a strong anti-correlation
between the isoscalar spin-orbit and tensor coupling constants. This
relationship is confirmed in our optimization through  a large
correlation coefficient of $-0.88$ between $C_{0}^{\rho\nabla J}$ and
$C_{0}^{JJ}$. In fact, the values of $C_{0}^{\rho\nabla J}$ and
$C_{0}^{JJ}$ are consistent with  the empirical
$C_{0}^{\rho\nabla J}(C_{0}^{JJ})$ dependence reported in
Ref.~\cite{(Les07)}. Yet, in spite of this very strong correlation, the value
of $C_{0}^{\rho\nabla J}$ changes only by 13\% between {\UNEDFONE} and
{\UNEDFTWO}.

\begin{table}[!htb]
\begin{center}
\caption{Comparison of parameter values for SLy4 and all three functionals
{\UNEDFZERO}, {\UNEDFONE}, and {\UNEDFTWO}.}
\label{table:unedfs}
\begin{ruledtabular}
\begin{tabular}{lcrrr}
$ \xb $                    & SLy4      & {\UNEDFZERO} & {\UNEDFONE} & {\UNEDFTWO} \\
\hline
$ \rhoc $                  &    0.16000 &    0.16053  &    0.15871  &    0.15631  \\
$ \enm/A $                 &  -15.972   &  -16.056    &  -15.8      &  -15.8      \\
$ \knm $                   &  229.901   &  230.0      &  220.0      &  239.930    \\
$ \asym $                  &   32.004   &   30.543    &   28.987    &   29.131    \\
$ \lsym $                  &   45.962   &   45.080    &   40.005    &   40.0      \\
$ 1/M_{s}^{*} $            &    1.439   &    0.9      &    0.992    &    1.074    \\
$ C_{0}^{\rho\Delta\rho} $ &  -76.996   &  -55.261    &  -45.135    &  -46.831    \\
$ C_{1}^{\rho\Delta\rho} $ &  +15.657   &  -55.623    & -145.382    & -113.164    \\
$ V_0^n $                  & -258.200   & -170.374    & -186.065    & -208.889    \\
$ V_0^p $                  & -258.200   & -199.202    & -206.580    & -230.330    \\
$ C_{0}^{\rho\nabla J}$    &  -92.250   &  -79.531    &  -74.026    &  -64.309    \\
$ C_{1}^{\rho\nabla J}$    &  -30.750   &   45.630    &  -35.658    &  -38.650    \\
$ C_{0}^{JJ}$              &    0.000   &    0.000    &    0.000    &  -54.433    \\
$ C_{1}^{JJ}$              &    0.000   &    0.000    &    0.000    &  -65.903    \\
\end{tabular}
\end{ruledtabular}
\end{center}
\end{table}

Looking more closely at the values of all spin-orbit and tensor coupling
constants of the {\UNEDFTWO} parameterization, we find that they are compatible
with the results of Ref.~\cite{(Zal08)}. In particular, the value of
$C_{0}^{\rho\nabla J}$ is close to the ``universal'' value of $-60$ MeV
obtained from a refit of three different EDFs to the spin-orbit splittings in
$^{40,48}$Ca and $^{56}$Ni. Our optimized tensor coupling constants,
as well as their isoscalar-isovector trend, are also in the same ballpark as
those partial refits of Ref.~\cite{(Zal08)}. They are, however, just beyond the space of
the T$IJ$ family of parameterizations considered in Ref.~\cite{(Les07)}. Recent
work also suggests that only the region where $C_{0}^{JJ} + C_{1}^{JJ} < 0$ and
$C_{0}^{JJ} - C_{1}^{JJ} > 0$ should be physical \cite{(Gra13)}. This is
the case for the {\UNEDFTWO} functional. Since the
{\UNEDFTWO} fit is an optimization carried out by considering a broad range of
nuclear properties (with five different types of experimental data), it is
encouraging that our results overlap well  with those obtained in
systematic studies of spin-orbit and tensor terms.

\begin{table}[!htb]
\begin{center}
\caption{{\UNEDFTWO} coupling constants in natural units. The value for the
scale is $\Lambda=687\,{\rm MeV}$.}
\label{table:natunit}
\begin{ruledtabular}
\begin{tabular}{ccccccc}
Channel & $C^{\rho\rho}_{t0}$ & $C^{\rho\rho}_{t{\rm D}}$ & $C^{\rho\tau}_{t}$ &
          $C^{\rho\Delta\rho}_{t}$ & $C^{\rho\nabla J}_{t}$ & $C^{JJ}_{t}$  \\
\hline
$t=0$ & $-$0.733 &  0.791 &  0.134 & $-$0.639 & $-$0.878 & $-$0.743 \\
$t=1$ &  0.328 & $-$0.291 & $-$0.319 & $-$1.545 & $-$0.528 & $-$0.900 \\
\end{tabular}
\end{ruledtabular}
\end{center}
\end{table}

We show in Table~\ref{table:natunit} the {\UNEDFTWO} coupling constants in
natural units \cite{(Fur97),(Kor10a)}. According to the hypothesis of
naturalness, the magnitude of (the absolute value of) coupling constants should
be of order unity, when scaled into unitless quantities. The scale $\Lambda$ used to perform the
transformation to natural units was taken as $\Lambda=687\,{\rm MeV}$, which
was found in Ref.~\cite{(Kor10a)} to be valid for Skyrme EDFs. As seen in
Table~\ref{table:natunit}, nearly all the {\UNEDFTWO} coupling constants fall
in the interval $[1/3, 3]$ which is compatible with the hypothesis of
naturalness \cite{(Kor10a)}. The one notable exception is $C^{\rho\tau}_{0}$,
which is unnaturally small; $C^{\rho\rho}_{1{\rm D}}$ and $C^{\rho\tau}_{1}$
are also at the limits of the allowed interval.


\subsection{Sensitivity Analysis}
\label{subsec:sensitivity}

The standard deviations $\sigma$ of the {\UNEDFTWO} parameterization are listed
in Table \ref{table:unedf2}, together with the 95\% confidence intervals. We
recall that the standard deviations (and also correlations) are calculated only
among those parameters that do not run into their imposed boundaries. Compared
with the previous parameterizations {\UNEDFZERO} and {\UNEDFONE}, the standard
deviations are overall smaller, reflecting improved constraints on the coupling
constants. For example, the standard deviation of the symmetry energy went down
from 3.05 MeV for {\UNEDFZERO} to 0.60 MeV for {\UNEDFONE} to only 0.32 MeV for
{\UNEDFTWO}. Similarly, the isoscalar effective mass, which could not be
constrained in  {\UNEDFZERO}, had a standard deviation of 0.12 for {\UNEDFONE},
which was further reduced to 0.05 for {\UNEDFTWO}. This improvement on
constraining all coupling constants of the functional, while not perfect, is a
confirmation of the validity of our strategy.

\begin{table*}[t!htb]
\begin{center}
\caption{Correlation matrix for the {\UNEDFTWO} parameter set. Absolute values
larger than 0.8 are printed in boldface.}
\label{table:corrMat}
\begin{ruledtabular}
\begin{tabular}{lcccccccccccc}
$ \rhoc $                &  1.00       &       &             &             &             &       &             &       &             &       &      & \\ 
$ \knm $                 & {\bf -0.97} &  1.00 &             &             &             &       &             &       &             &       &      & \\
$ \asym $                & -0.07       & -0.03 &  1.00       &             &             &       &             &       &             &       &      & \\ 
$ 1/M_{s}^{*} $          &  0.08       & -0.05 & -0.24       &  1.00       &             &       &             &       &             &       &      & \\ 
$C_{0}^{\rho\Delta\rho}$ & -0.43       &  0.43 &  0.22       & {\bf -0.89} &  1.00       &       &             &       &             &       &      & \\ 
$C_{1}^{\rho\Delta\rho}$ & -0.42       &  0.37 &  {\bf 0.83} & -0.17       &  0.31       &  1.00 &             &       &             &       &      & \\ 
$ V_0^n $                & -0.06       &  0.02 &  0.27       & {\bf -0.96} &  {\bf 0.85} &  0.17 &  1.00       &       &             &       &      & \\ 
$ V_0^p $                & -0.09       &  0.05 &  0.21       & {\bf -0.89} &  {\bf 0.80} &  0.14 &  {\bf 0.86} &  1.00 &             &       &      & \\ 
$ C_{0}^{\rho\nabla J}$  & -0.51       &  0.50 &  0.34       & -0.40       &  0.68       &  0.55 &  0.36       &  0.34 &  1.00       &       &      & \\ 
$ C_{1}^{\rho\nabla J}$  & -0.31       &  0.29 & -0.19       & -0.00       &  0.04       &  0.18 & -0.07       & -0.02 &  0.14       &  1.00 &      & \\ 
$ C_{0}^{JJ}$         &  0.56       & -0.55 & -0.26       &  0.05       & -0.35       & -0.53 & -0.02       & -0.02 & {\bf -0.88} & -0.35 & 1.00 & \\ 
$ C_{1}^{JJ}$         &  0.36       & -0.35 &  0.13       & -0.23       &  0.16       & -0.14 &  0.29       &  0.25 & -0.02       & -0.57 & 0.29 & 1.00  \\
\hline
 & $ \rhoc $ & $ \knm $  & $ \asym $ & $ 1/M_{s}^{*} $ & $ C_{0}^{\rho\Delta\rho} $ & $ C_{1}^{\rho\Delta\rho} $ &
   $ V_0^n $ & $ V_0^p $ & $ C_{0}^{\rho\nabla J}$ & $ C_{1}^{\rho\nabla J}$ & $ C_{0}^{JJ}$ & $ C_{1}^{JJ}$ \\
\end{tabular}
\end{ruledtabular}
\end{center}
\end{table*}

Table \ref{table:corrMat} displays the correlation matrix among the coupling
constants of the functional. As with {\UNEDFONE}, there exists a strong
correlation between the pairing strength parameters and both the isoscalar
effective mass and the isoscalar surface coefficient $C_{0}^{\rho\Delta\rho}$,
although for the latter the correlation is less pronounced than for
{\UNEDFONE}. These correlations reflect a strong interplay between the level
density near the Fermi level and the magnitude of pairing correlations. Another
strong (anti-)correlation can be observed between  $\rhoc$ and $\knm$. This is
reminiscent of our first parameterization, dubbed {\UNEDFNB} in
Ref.~\cite{(Kor10)}, where no bounds had been imposed on the coupling
constants. Further, as discussed in Sec.~\ref{subsec:UNEDF2}, the isoscalar
spin-orbit and tensor coupling constants are strongly anti-correlated and seem
to follow the  trend predicted in Ref.~\cite{(Les07)}.

\begin{figure}[!htb]
\includegraphics[width=\linewidth]{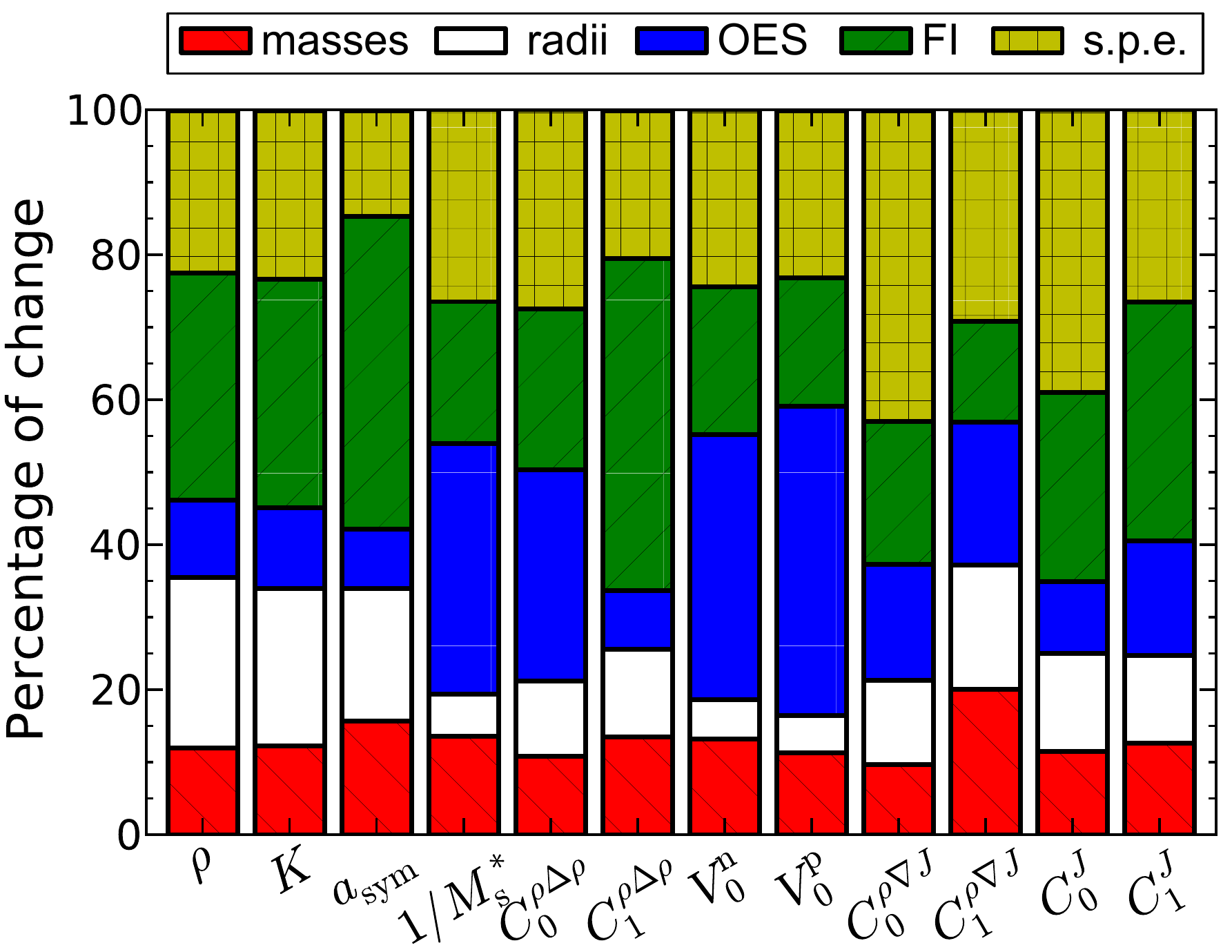}
\caption{(Color online) Sensitivity of the {\UNEDFTWO} parameterization to
different data types: masses, charge radii, OES, fission isomer excitation
energies (FI), and s.p.\ energies.}
\label{fig:sensparam}
\end{figure}

The overall impact of each data type on the {\UNEDFTWO}
parameterization can be assessed by studying the sensitivity matrix $S$:
\begin{equation}
S(\xb) = \left[ J(\xb)J^{\rm T}(\xb) \right]^{-1}J(\xb) \, ,
\end{equation}
where $J(\xb)$ is the Jacobian matrix calculated with the parameterization
$\xb$. Following Refs.~\cite{(Kor10),(Kor12)}, we have calculated the partial
sums of the absolute values in $S(\xb)$ for each data type, normalized with
respect to the number of data points. The results are presented in
Fig.~\ref{fig:sensparam}, with each bar normalized to 100\%. A number of
observations made for the {\UNEDFZERO} or {\UNEDFONE} functionals still apply,
such as the strong sensitivity of pairing strengths to OES data or the large
impact of fission isomer excitation energies on the determination of $\asym$.
Overall, s.p.\ splittings, fission isomer excitation energies, and OES data
seem to be the main drivers of the parameterization, while the relative role of
masses is reduced. Looking closely at the coupling constants that are
relatively well constrained, one may identify two trends: (i) bulk coupling
constants (i.e., $\rho$, $\knm$, and $\asym$) are not really impacted by the
OES data; (ii) surface coupling constants (involving gradient terms) are more
sensitive to OES data, fission isomer excitation energies, and s.p.\ 
splittings. The three isovector surface coupling constants
($C_{1}^{\rho\Delta\rho}$, $C_{1}^{\rho\nabla J}$, $C_{1}^{JJ}$) behave
differently but are less constrained by the data, as shown by their large
standard deviations reported in Table \ref{table:unedf2}.

\begin{figure}[!htb]
\center
\includegraphics[width=\linewidth]{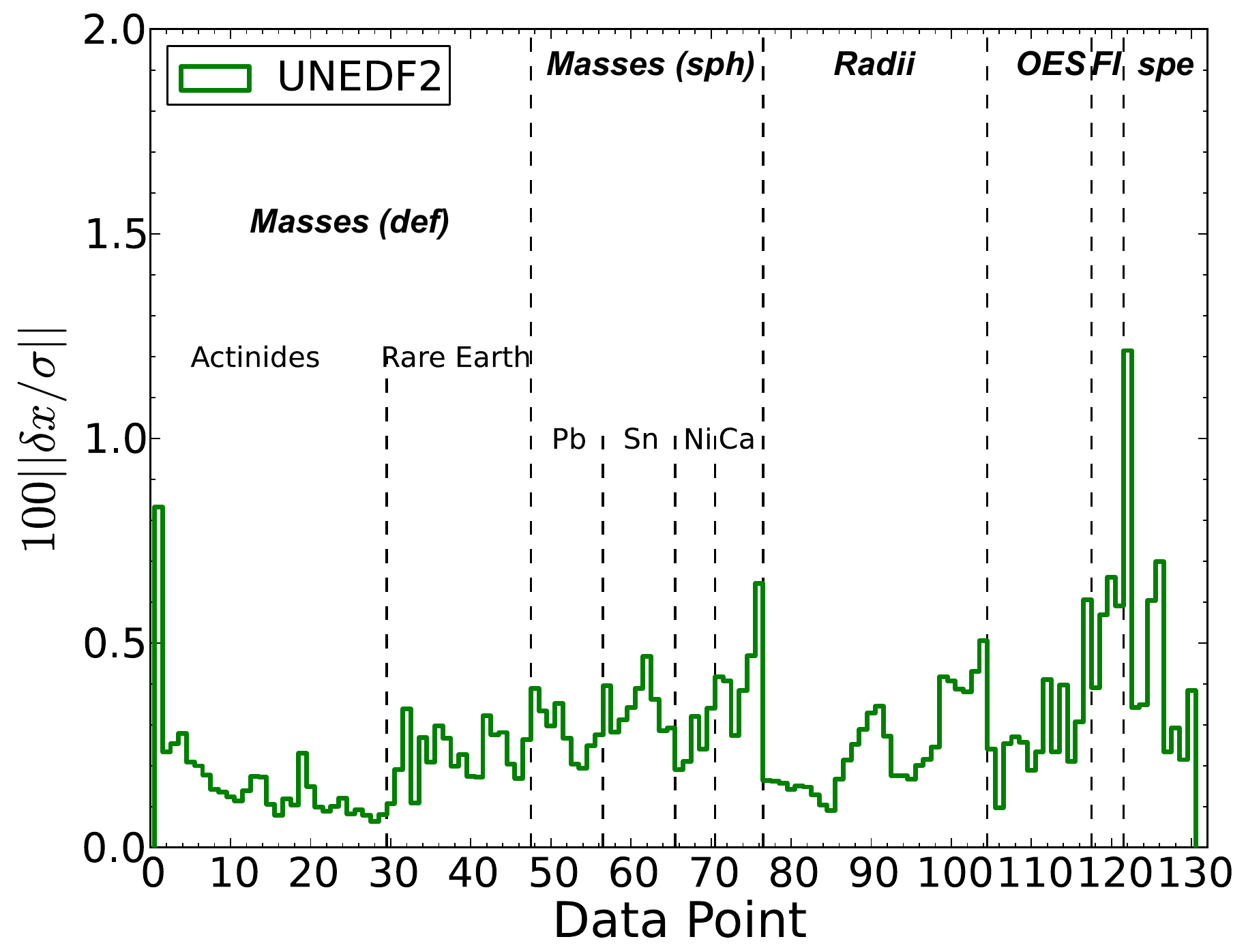}
\caption{(Color online) Overall change of the {\UNEDFTWO} parameterization when
data point $d_{i,j}$ is changed (in isolation) by $0.1\,w_{i}$.}
\label{fig:sensdata}
\end{figure}

A complementary way to examine our optimization dataset is to analyze the
impact of individual data points on the optimized solution. This is plotted in
Fig~\ref{fig:sensdata}. Here, the amount of variation
\begin{equation}
\vert\vert {\delta \xb} / {\sigma} \vert\vert
= \sqrt{\sum_{k=1}^{n_{x}}\left( \frac{\delta x_{k}}{\sigma_{k}} \right)^{2}}
\end{equation}
for the optimal solution is presented when each data point $d_{i,j}$ is shifted
by an amount of $0.1\,w_{i}$. The standard deviations $\sigma_{k}$ for
parameters $x_{k}$ are listed in Table \ref{table:unedf2}. As for {\UNEDFZERO}
and {\UNEDFONE}, the overall changes in $\xh$ are of the same order of
magnitude and are very small, $\| \delta\gras{x}/\sigma \|\approx 0.01$.
This indicates that the  set of fit observables in {\UNEDFTWO} has been chosen
consistently. The new s.p.\ data points seem to have a relatively large impact
on the parameterization, especially the s.p.\ splittings in $^{40}$Ca.


\section{Properties of UNEDF2 parameterization}
\label{sec:results}

In this section, we review various properties of the {\UNEDFTWO}
parameterization. In Sec.~\ref{subsec:linearresponse}, we apply the linear
response theory to test the functional against the presence of finite-size
instabilities. Section~\ref{subsec:correlations} discusses correlations among
various  observables. Predictions of {\UNEDFTWO} for shell structure in
doubly-magic nuclei are presented in Sec.~\ref{subsec:shell}, and global
binding energy and deformation trends (in particular in the context of nuclear
fission) are analyzed in Secs.~\ref{subsec:masses} and \ref{subsec:fission}, respectively. Section~\ref{subsec:droplets} contains
the discussion of {\UNEDFTWO} predictions for neutron droplets in external
traps.


\subsection{Linear Response and Instabilities}
\label{subsec:linearresponse}

The linear response formalism in nuclear physics has been developed mainly
in the framework of the Random Phase Approximation (RPA) based on the use of
an effective interaction. In Refs.~\cite{(Pas12a),*(Pas12b)} this formalism was
generalized to determine the response function in both symmetric nuclear matter and pure
neutron matter for the case of a general Skyrme EDF as given in
Ref.~\cite{(Per04)}. This recent development is needed here because of the
non-standard spin-orbit term of the {\UNEDF} parametrizations. The response functions $\chi^{(S,M,T)}(\omega,\mathbf{q})$ of
interest to the present work are defined as the response of the infinite medium
to external probes of the type
\begin{equation}
\hat{Q}^{(S,M,T)}=\sum_{j}e^{i\mathbf{q}\cdot
\mathbf{r}_{j}} \Theta_{j}^{(S,M,T)} \, ,
\end{equation}
where $S$ ($M$) is the spin (its projection along the $z$-axis), $T$ is the
isospin, and $\Theta_{j}^{(S,M,T)}$ is an operator acting on spin, isospin, or
both. As recently shown in Ref.~\cite{(Pas12c)}, the response function of
neutron matter can provide information about instabilities in finite
nuclei \cite{(Les06)}. More precisely, it was shown that whenever a pole
appears in the response function close enough to the saturation density of the
system $\rho_{\rm c}$, the finite nucleus undergoes an instability in the
corresponding channel (scalar/vector, isoscalar/isovector), see examples in
Ref.~\cite{(Sch10)}.

Several quantitative criteria to estimate the likelihood of finite-size
instabilities for a given EDF have been recently proposed \cite{(Pas12d),(Hel13)}.
Because of shell fluctuations, the nucleus can explore regions of densities
slightly larger than the saturation density. In Ref.~\cite{(Hel13)}, the
following conservative criterion was established: whenever the response function has a
pole at a density $\rho\approx1.4\rho_{\rm c}$, there is a risk of instability
in calculations for finite nuclei. More complex criteria were proposed in \cite{(Pas12b)}. The poles of the response function are in
practice determined by solving the equation
\begin{equation}
1/\chi^{(S,M,T)}(\omega=0,\mathbf{q})=0,
\end{equation}
where the expressions of $\chi^{(S,M,T)}$ can be found in Ref.~\cite{(Pas12a)}.

\begin{figure}[!htb]
\includegraphics[width=\linewidth]{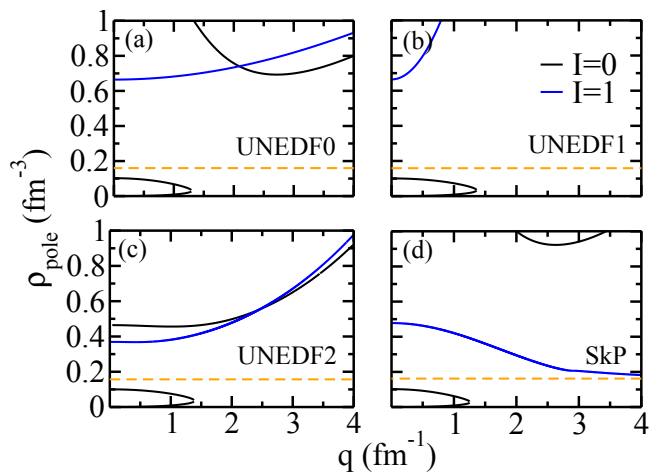}
\caption{(Color online) Position of the lowest critical densities for
$S=0,\, M=0$. Solid line: isoscalar $T=0$ channel; dashed line: isovector $T=1$
channel. The horizontal dashed line shows the saturation density.}
\label{fig:unedfS0}
\end{figure}

Figure~\ref{fig:unedfS0} shows the position of the lowest poles of the response
function as a function of the transferred momentum $\gras{q}$ in symmetric
nuclear matter for {\UNEDFZERO}, {\UNEDFONE}, {\UNEDFTWO}, and SkP
\cite{(Dob84)}. Since the {\UNEDF} functionals have been developed to be used
only in the time-even channel, we will limit our analysis to $S=0$. We first
remark that in all cases we have an instability in the region at low density
and low momentum in the channel $S=0,\, M=0,\, T=0$. This is the well-known
spinodal instability, which is physical. We then observe that all {\UNEDF}
functionals satisfy the stability criterion given in Ref.~\cite{(Hel13)}. On
the other hand, SkP is unstable in the scalar/isovector channel in the region
of densities around $\rho_{c}$; as a result it was shown in Ref.~\cite{(Les06)}
that calculations of finite nuclei with this functional are unstable.

\begin{figure}[!htb]
\includegraphics[width=\linewidth]{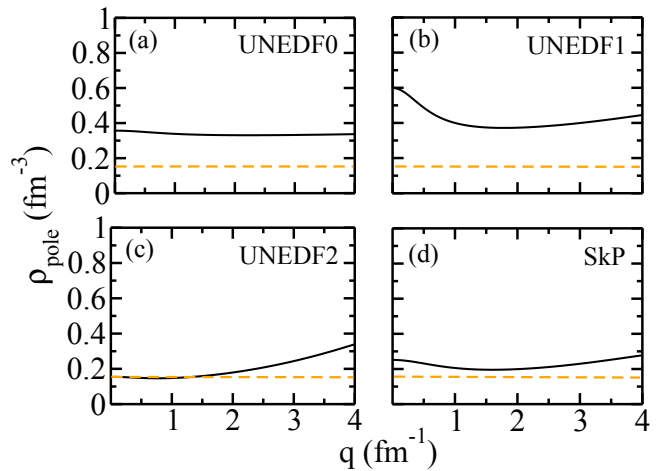}
\caption{Position of the lowest critical densities for $S=0$ in pure
neutron matter. The horizontal dashed line shows the saturation density.}
\label{fig:pnm}
\end{figure}

Although a quantitative criterion is not yet available for pure neutron matter,
it is interesting  to study the instabilities of such a system.
Figure~\ref{fig:pnm} shows the position of the lowest poles for pure neutron
matter. While the {\UNEDFZERO}, {\UNEDFONE}, and (to a lesser extent) SkP EDFs do
not exhibit poles near the saturation density, the situation is different for
{\UNEDFTWO}. This suggests that $S=0$ instabilities could manifest in
neutron-rich nuclei or in trapped neutron droplets. As we will see in
Sec.~\ref{subsec:droplets} below, such instabilities are indeed present in
heavy neutron droplets calculated with {\UNEDFTWO}. This result may be a
consequence of the large negative values of the tensor coupling constants
\cite{(Les07)}.


\subsection{Correlations with Other Observables}
\label{subsec:correlations}

The sensitivity analysis presented in Sec.~\ref{subsec:sensitivity} aimed at
quantifying the behavior of the $\chi^{2}$ landscape at the minimum for the set
of observables used in the fit. Complementary information can be obtained from
an analysis of the correlations between observables {\em not} included in the
fit \cite{(Rei10),(Gao13),(Kor13)}. These correlations can be extracted from an
estimate of confidence regions near the minimum. In this section, we define the
confidence region based on a criterion for the value of the objective function
\cite{(Bra99)}. It is a slightly different prescription from the procedure that
we use to define the confidence intervals; see Sec. III.B.1. of
Ref.~\cite{(Kor10)} for details. Asymptotically (at the limit of
$n_{d}\rightarrow +\infty$), both prescriptions are in fact equivalent (see
discussion in Sec. 3.3.1 of Ref.~\cite{(NLR89)}).

\begin{table}[!htb]
\caption{Calculated values and standard deviation $\sigma$ for various
observables computed with {\UNEDFTWO}: saturation density $\rho_{\rm c}$ (in
fm$^{-3}$); incompressibility of symmetric nuclear matter $K$ (in MeV);
isoscalar effective mass $M_s$; symmetry energy ${\asym}$ (in MeV); slope of
the neutron equation-of-state $d_{\rho} (E/N)$ at $\rho=\rho_{\rm c}/2$ (in
MeV\,fm$^3$); peak energies of giant resonances in  $^{208}$Pb (isoscalar
monopole, GMR; isoscalar quadrupole, GQR; isovector dipole; GDR; all in MeV);
electric dipole polarizability $\alpha_D$  in  $^{208}$Pb (in fm$^2$/MeV); and
neutron skin $r_n-r_p$ in $^{208}$Pb (in fm).
}
\label{table:obsvariation}
\begin{ruledtabular}
\begin{tabular}{lcc}
 Observable & $\overline{A}$  & $\sigma_A$ \\
\hline
$\rho_c$          &     0.156 &  0.001 \\
$K$               &   240     & 10     \\
$M_s^*$           &     0.93  &  0.04  \\
${\asym}$         &    29.1   &  0.3   \\
$d_{\rho_n}(E/N)$ &    75.8   &  2.2   \\ 
GMR               &    14.0   &  0.3   \\
GQR               &    10.8   &  0.3   \\
GDR               &    13.6   &  0.1   \\
$\alpha_D$        &    13.8   &  0.1   \\
$r_n-r_p$         &     0.167 &  0.003
\end{tabular}
\end{ruledtabular}
\label{table:correlation}
\end{table}

Here we construct an approximate confidence region using the following
approach. At the minimum $\gras{x}_{\text{fin}}$, the quantity
$\chi_0=\chi^{2}(\gras{x}_{\text{fin}})$ characterizes the best-fit
parameterization. The parameters $\gras{x}$ in a neighborhood $\mathcal{V}$ of
the minimum can still provide a reasonable description of nuclear properties.
The parameter space $\mathcal{V}$ is thus referred to as a ``reasonable" domain.
Each observable $A$ that can be computed in the EDF theory is also a function
of the Skyrme parameters, $A=A(\gras{x})$. Varying $\gras{x}$ in the vicinity
of the optimal set will lead to fluctuations in the values of $A$ with respect
to its value at the minimum, $A_{0} = A(\gras{x}_{\text{fin}})$. The
uncertainty of the prediction is characterized by the variance
$\text{Var}(A)=\sigma_A^2=\overline{(A - A_{0})^{2}}$, where the average value
is computed from
\begin{equation}
\overline{A}=\int_{\mathcal{V}} d\gras{x}\,W(\gras{x})A(\gras{x}).
\label{eq:average}
\end{equation}
This simple estimate of uncertainties provides valuable information on the
predictive power of the model. Further information can be obtained from the
correlation coefficient $c_{AB}$ between two observables $A$ and $B$
defined from the covariance matrix as
\begin{equation}\label{cab}
c_{AB} = \frac{\text{Cov}(A,B)}{\sqrt{\text{Var}(A)\text{Var}(B)}}.
\end{equation}

\begin{figure}[!htb]
\includegraphics[width=\linewidth]{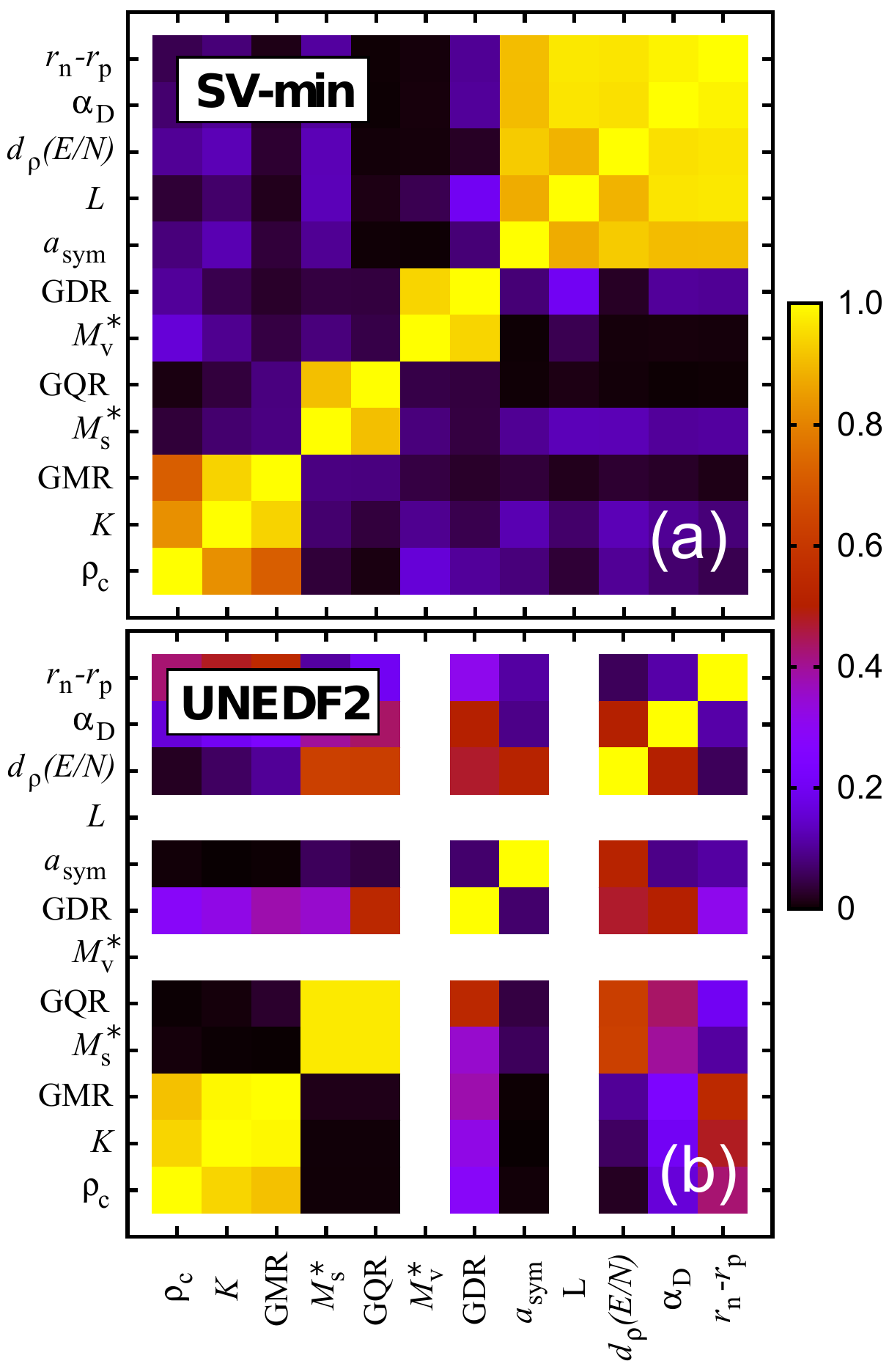}
\caption{(Color online) Absolute values of correlations between various
observables for (a) SV-min and (b) {\UNEDFTWO}.
The expectation values of individual observables and their uncertainties are
listed in Table~\ref{table:obsvariation}.
  }
\label{fig:obscorrelations}
\end{figure}

Table \ref{table:correlation} shows the values and the uncertainties of a large
set of observables. The vicinity $\mathcal{V}$ was defined by the level set
$\mathcal{V}= \left\{ \gras{x}: \chi^{2}(\gras{x})- \chi_{0} \leq n_{\rm d}-n_{\rm x}\right\}$.
As noted earlier, this construction of $\mathcal{V}$ is different from the one
used to define the 95\% confidence interval, and hence, the standard deviations
reported in Table~\ref{table:correlation} are slightly different from those of
Table~\ref{table:unedf2}. As discussed later in Sec.~\ref{subsec:sensitivity}, the set {\UNEDFTWO} delivers rather small
uncertainties for all observables shown.

Figure~\ref{fig:obscorrelations} shows the correlation matrix $c_{AB}$
(\ref{cab}) between various pairs of observables computed for $^{208}$Pb. To
illustrate the impact of the optimization protocol, we compare {\UNEDFTWO} with
the SV-min parameterization \cite{(Klu09)}. In the case of SV-min (upper
panel), we can see four blocks of highly correlated observables
\cite{(Rei13),(Pie12),(Naz13)}: (i) the nuclear incompressibility $K$ with
saturation density $\rho_c$ and the peak of the giant monopole resonance; (ii)
the isoscalar effective mass $M_s^{*}$ with the peak of the giant quadrupole
resonance; (iii) the isovector effective mass $M_v^{*}$ with the peak of the
giant dipole resonance; and (iv) a  block of correlated isovector indicators
\cite{(Rei10)}: $a_\mathrm{sym}$, $L$, $d_{\rho_n}(E/N)$, $\alpha_D$, and
$r_n-r_p$.

\begin{figure}[!htb]
\includegraphics[width=\linewidth]{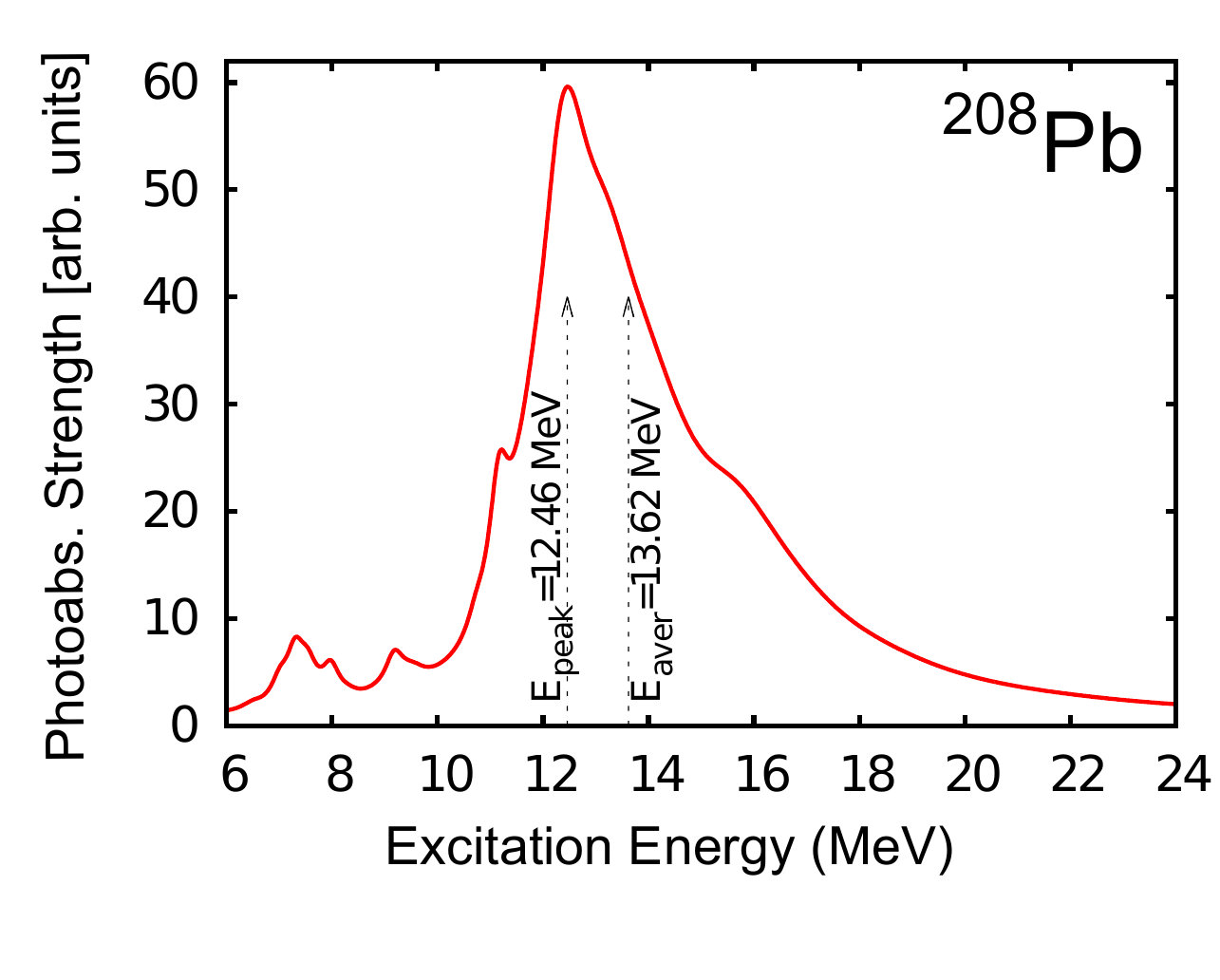}
\caption{(Color online) The E1 strength distribution for $^{208}$Pb
computed with {\UNEDFTWO}.}
\label{fig:GDR}
\end{figure}

As seen in Fig.~\ref{fig:obscorrelations}, {\UNEDFTWO} is missing some of the
correlations predicted by SV-min. The reason is essentially that {\UNEDFTWO}
has three symmetric neutron matter parameters fixed. Two of them, the isovector
effective mass $M_v^*$ and the slope of symmetry energy $L$, constitute crucial
constraints because they are related to the properties of the linear response.
Since they are fixed, they have been eliminated from the correlation matrix
(white rows and columns). This step leaves the peak of the GDR unconstrained
but constrains $\asym$ considerably (uncertainty of 0.3\,MeV for {\UNEDFTWO},
compared with 1.7\,MeV for SV-min). Consequently, nearly all correlations
between $\asym$ and the isovector static observables have disappeared. Another
consequence of freezing $\lsym$ is the relatively small uncertainty for the
neutron skin. These are due to the fact that the largest contribution to the
error budget of $r_n$ comes from $\lsym$ \cite{(Kor13)}.

The {\UNEDFTWO} value of $\asym$ is consistent with the current estimates
\cite{(Lat12),(Naz13)}, and the same holds for $\alpha_D$ and $r_n-r_p$
\cite{(Rei10),(Pie12),(Rei13),(Rei13a)}. A word of caution is in order
concerning the peak position of the GDR. The energies of giant resonance peaks
given in Table \ref{table:obsvariation} stem from an average over a broad
energy region. Figure~\ref{fig:GDR} shows the detailed energy-weighted dipole
strength computed for $^{208}$Pb  with {\UNEDFTWO}. The RPA results are folded
with an energy-dependent width in order to simulate the increase of collisional
width with excitation energy \cite{(Rei13)}. The GDR peak is strongly
fragmented because it resides in a region of large density of $1ph$ states. The
$1ph$ fragmentation is asymmetric because the density of $1ph$ states increases
with energy. This produces a discrepancy between the averaged excitation energy
$E_\mathrm{aver}$ (the average taken just over the resonance region by virtue
of a fluid dynamics approach) and the peak energy $E_\mathrm{peak}$, which is
considerably smaller. The experimental energy of the GDR resonance,
13.6\,MeV, corresponds to the peak energy. Consequently, we find that
{\UNEDFTWO}, similar to SV-min and several other Skyrme EDFs, underestimates
the GDR peak energy. To overcome this problem, a smaller isovector effective
mass or larger TRK sum rule enhancement is required \cite{(Klu09),(Rei13)}.


\subsection{Shell Structure}
\label{subsec:shell}

One of the primary motivations behind this work was to use experimental data on
s.p.\ splittings to optimize the tensor coupling constants of the Skyrme EDF.
In Table~\ref{table:shellgap}, we report the root-mean-square deviations from
experimental data for binding energies for 24 odd-$A$ nuclei that are one mass
unit away from the doubly magic systems $^{16}$O, $^{40,48}$Ca, $^{56}$Ni,
$^{132}$Sn, and $^{208}$Pb. (In the following, the abbreviation RMSD will
always stand for a root-mean-square deviation between theoretical values and
experimental data or empirical estimates.) The table also shows the RMSD for
(six) two-neutron and two-proton separation energies across each shell gap. For
example, in the case of $^{208}$Pb: $B(A_{\rm mag}-1)$ would stand for
$B(^{207}\mathrm{Tl})$ (protons) and $B(^{207}\mathrm{Pb})$ (neutrons).
Similarly, $B(A_{\rm mag}+1) \equiv B(^{209}\mathrm{Bi})$ for protons and
$B(A_{\rm mag}+1) \equiv B(^{209}\mathrm{Pb})$ for neutrons; $S_{2n}$
represents the two-neutron separation energy of $^{209}$Pb:
$S_{2n} \equiv B(^{209}\mathrm{Pb}) - B(^{207}\mathrm{Pb})$; and $S_{2p}$ is
the two-proton separation energy of $^{209}$Bi
$S_{2p} \equiv B(^{209}\mathrm{Bi}) - B(^{207}\mathrm{Tl})$. Since s.p.\ 
splittings are computed from binding energy differences of the neighboring
odd-$A$ nuclei, all these RMSDs are indicators of the quality of the underlying
single-particle spectra.

\begin{table}[htb]
\caption{Root-mean-square deviation from experiment for observables predicted
with SLy4,  {\UNEDFZERO}, {\UNEDFONE}, and {\UNEDFTWO} that are related to
magic gaps: binding energies $B(A_{\rm mag}\pm 1)$ of one-particle or one-hole
nuclei outside doubly magic systems, and  $S_{2n}$  and $S_{2p}$ values across
the shell gap (all in MeV). See text for details.}
\label{table:shellgap}
\begin{ruledtabular}
\begin{tabular}{lcccc}
Quantity & SLy4 & {\UNEDFZERO} & {\UNEDFONE} & {\UNEDFTWO} \\
\hline
$B(A-1)$               & 3.30 & 2.70 & 2.78 & 2.16 \\ 
$B(A+1)$               & 3.06 & 2.44 & 2.04 & 2.12 \\
$S_{2n}$ and $S_{2p}$  & 0.90 & 1.44 & 1.59 & 0.90
\end{tabular}
\end{ruledtabular}
\end{table}

One can see in the table that although {\UNEDFZERO} and {\UNEDFONE} reproduce
binding energies $B(A_{\rm mag}\pm 1)$ better than SLy4, the latter works
better for two-particle separation energies. The {\UNEDFTWO} parameterization
brings a significant improvement on binding energies with respect to
{\UNEDFZERO} while maintaining a decent reproduction of two-particle separation
energies. In spite of this progress, the resulting RMSDs are quite appreciable.

\begin{figure}[!htb]
\includegraphics[width=0.85\linewidth]{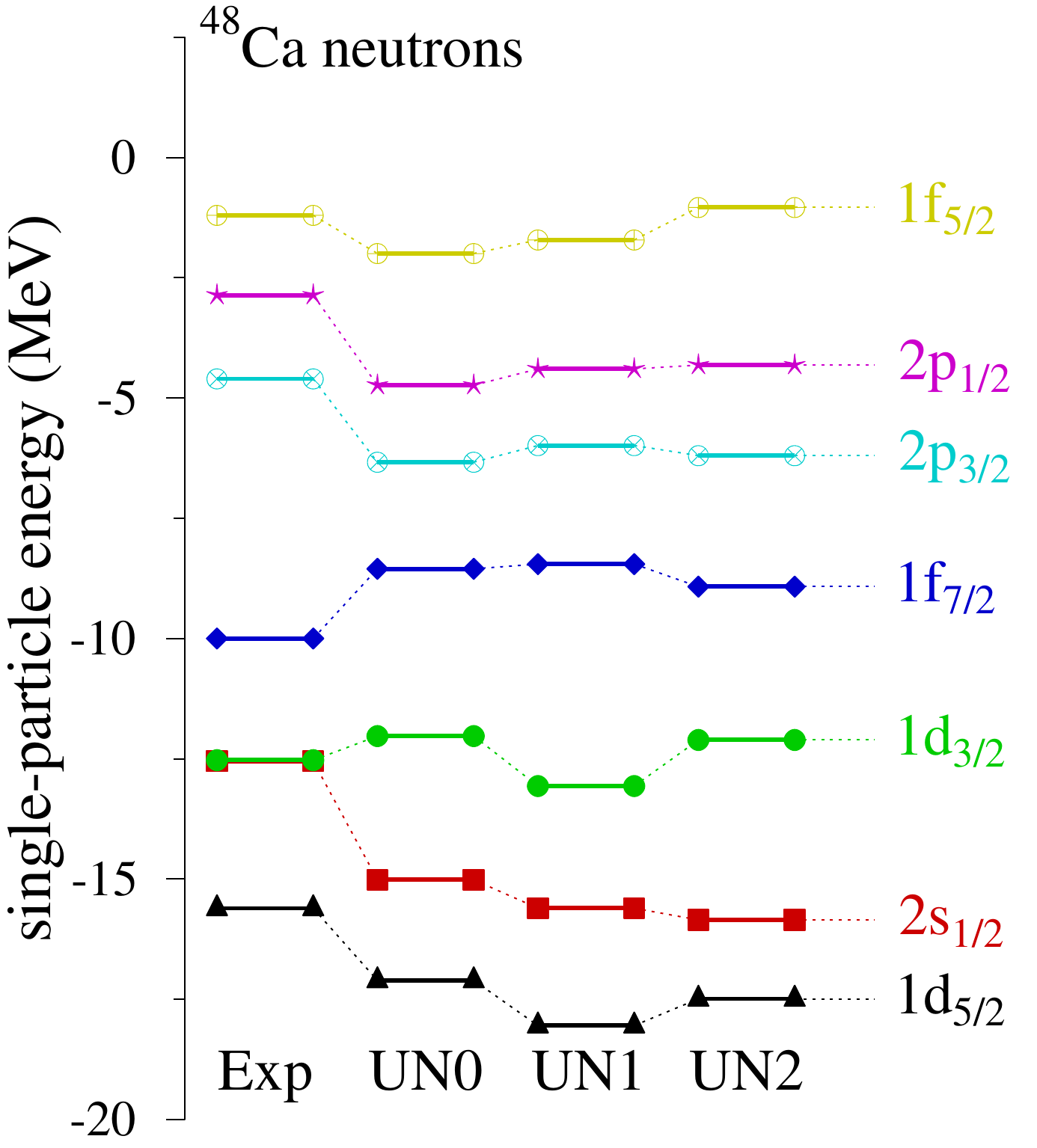}
\caption{(Color online) Neutron single-particle energies in $^{48}$Ca
calculated with the {\UNEDFZERO} (UN0), {\UNEDFONE} (UN1), and {\UNEDFTWO}
(UN2) parameterizations of the Skyrme energy density. These are compared with
the empirical values (Exp) of Ref.~\cite{(Sch07)}.}
\label{fig:spe48n}
\end{figure}

\begin{figure}[!htb]
\includegraphics[width=0.85\linewidth]{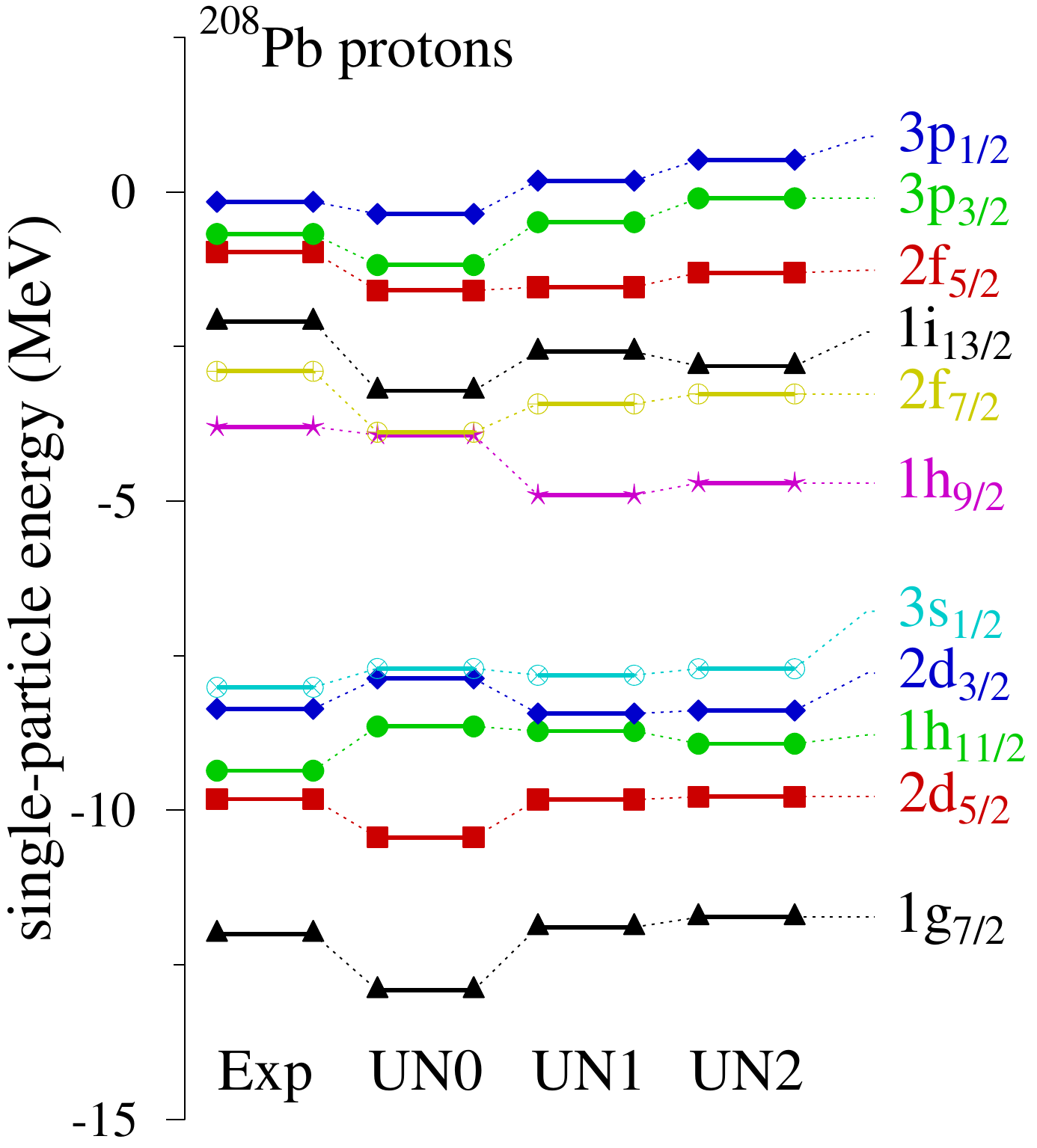}
\caption{(Color online) Same as Fig.~\ref{fig:spe48n} but for proton
single-particle energies in $^{208}$Pb.}
\label{fig:spe208p}
\end{figure}

\begin{figure}[!htb]
\includegraphics[width=0.85\linewidth]{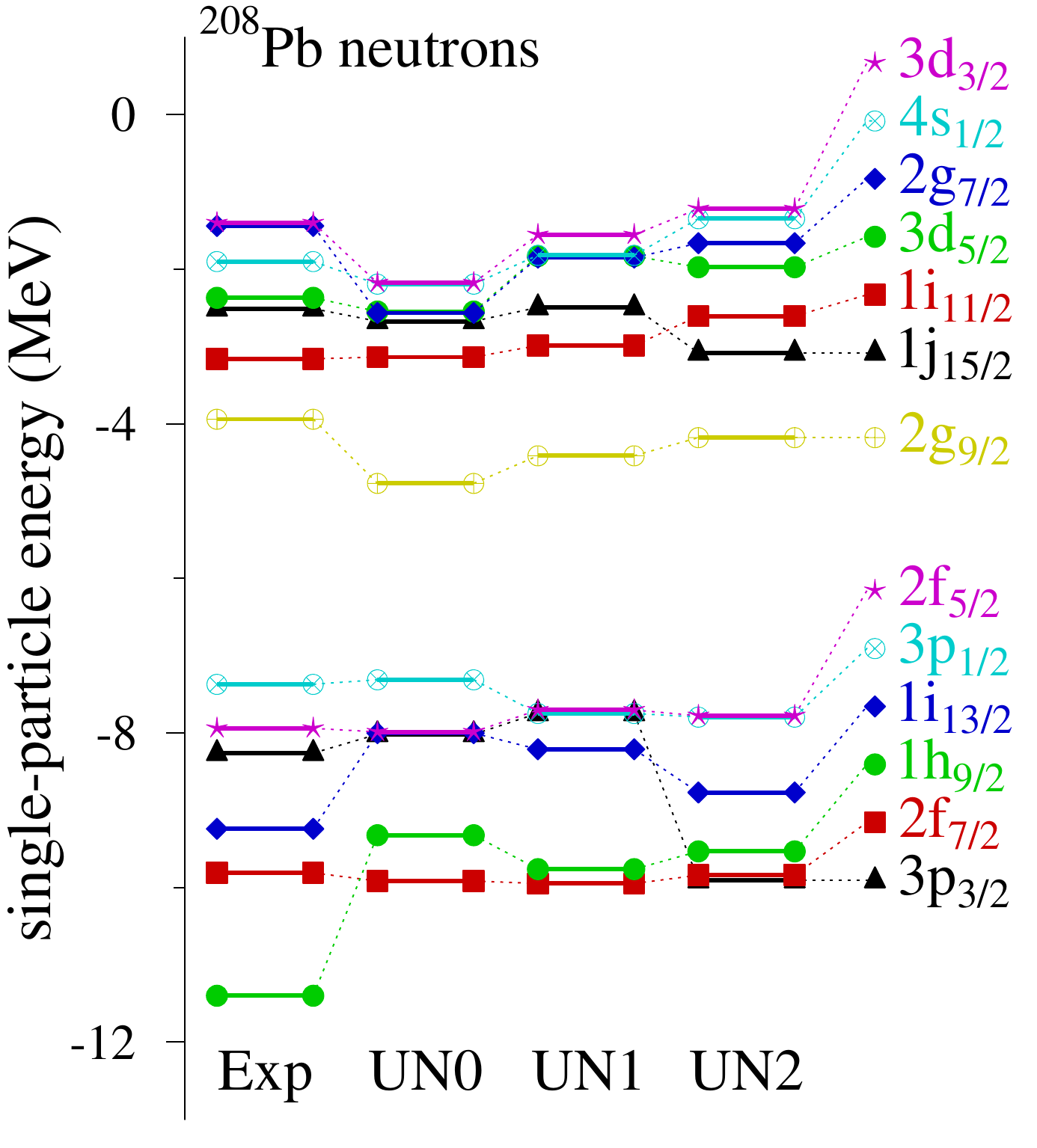}
\caption{(Color online) Same as Fig.~\ref{fig:spe208p} but for neutron
single-particle energies in $^{208}$Pb.}
\label{fig:spe208n}
\end{figure}

Figures~\ref{fig:spe48n}--\ref{fig:spe208n} display the s.p.\ levels, as defined by Eqs.(\ref{eq:spEpart})-(\ref{eq:spEhole}), for
neutrons in $^{48}$Ca and for protons and neutrons in $^{208}$Pb, respectively.
Compared with the empirical values, the $N=28$ gap  in $^{48}$Ca is clearly too
small with {\UNEDFTWO}. Otherwise, the positions of most of the levels seem to
be slightly improved compared with {\UNEDFONE}, which was itself a minor
improvement over {\UNEDFZERO}. The single-particle proton levels in $^{208}$Pb  show
that the $Z=82$ magic gap is also too small in {\UNEDFONE} and {\UNEDFTWO},
because of a low energy of the $h_{9/2}$ shell.  Further, we notice in
Fig.~\ref{fig:spe208n} the inversion of the 1$j_{15/2}$ and 1$i_{11/2}$ shells
and a large shift in the energy of 3$p_{3/2}$ shell. The spectra shown in
Figs.~\ref{fig:spe48n}-\ref{fig:spe208n} are quite representative of the
predictive power of the {\UNEDF} family with respect to shell structure.

\begin{table}[!ht]
\caption{RMSDs of s.p.\ energies from empirical values of Ref.~\cite{(Sch07)}
(in MeV).}
\label{table:rms_spenergy}
\begin{ruledtabular}
\begin{tabular}{lccc}
Nuclei & {\UNEDFZERO} & {\UNEDFONE} & {\UNEDFTWO} \\
\hline
All    & 1.42 & 1.38 & 1.38 \\ 
Light  & 1.80 & 1.72 & 1.74 \\
Heavy  & 0.94 & 0.97 & 0.95 \\
\end{tabular}
\end{ruledtabular}
\end{table}

To quantify further the quality of the predicted  shell structure, we list in
Table~\ref{table:rms_spenergy} the RMSDs of single-particle energies from the
empirical values of Ref.~\cite{(Sch07)}. The calculation is based on 75
(negative-energy) levels in the same set of double-magic nuclei as in Table
\ref{table:shellgap}. We have also partitioned the set of nuclei into light
($A<80$; 36 levels) and heavy nuclei ($A\ge 80$; 39 levels). Note that all
s.p.\ states used to compute the RMSDs were obtained from HFB calculations with
the blocking procedure.

Overall, the RMSD from experiment  is similar for all {\UNEDF}
parameterizations. The larger RMSD obtained for light nuclei is explained
mostly by a lower level density, which increases the average error. Also, the
impact of correlations missing in the Skyrme EFT approach is greater in lighter
systems, the structure of which is profoundly impacted  by surface effects.
Even though two-particle separation energies across the shell gap are improved
with {\UNEDFTWO}, the overall reproduction of shell structure is not.

These results are consistent with the conclusions of Ref.~\cite{(Kor08)}, where
it was found that Skyrme EDFs are intrinsically limited in their ability to
reproduce s.p.\ spectra in doubly-magic nuclei. The regression analysis
technique employed therein suggests that the best possible RMSD for s.p.\ 
energies obtained in the Skyrme EDF approach is around 1.2\,MeV. Although the
calculations of Ref.~\cite{(Kor08)} were performed at the HF level, it is
unlikely that using the physically better motivated blocking procedure, and
considering particle-vibration-coupling and self-interaction corrections
\cite{(Tar13)} would significantly alter the conclusions. The RMSD of 1.38 MeV
found for {\UNEDFTWO} is thus very close to the limit given by the regression
analysis, especially considering the diversity of constraints imposed during
the fit.


\begin{figure}[!htb]
\includegraphics[width=0.9\linewidth]{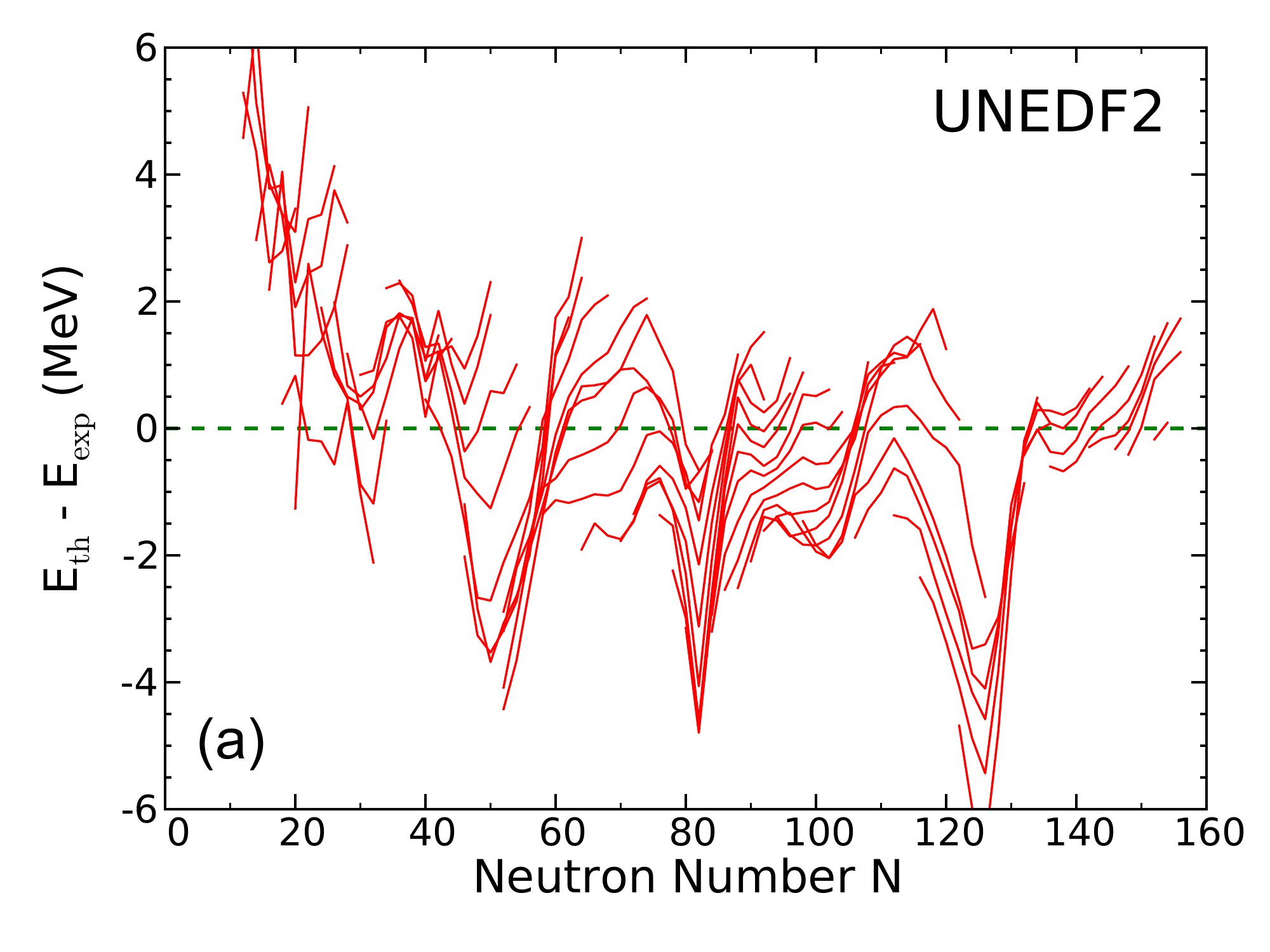} \\
\includegraphics[width=0.9\linewidth]{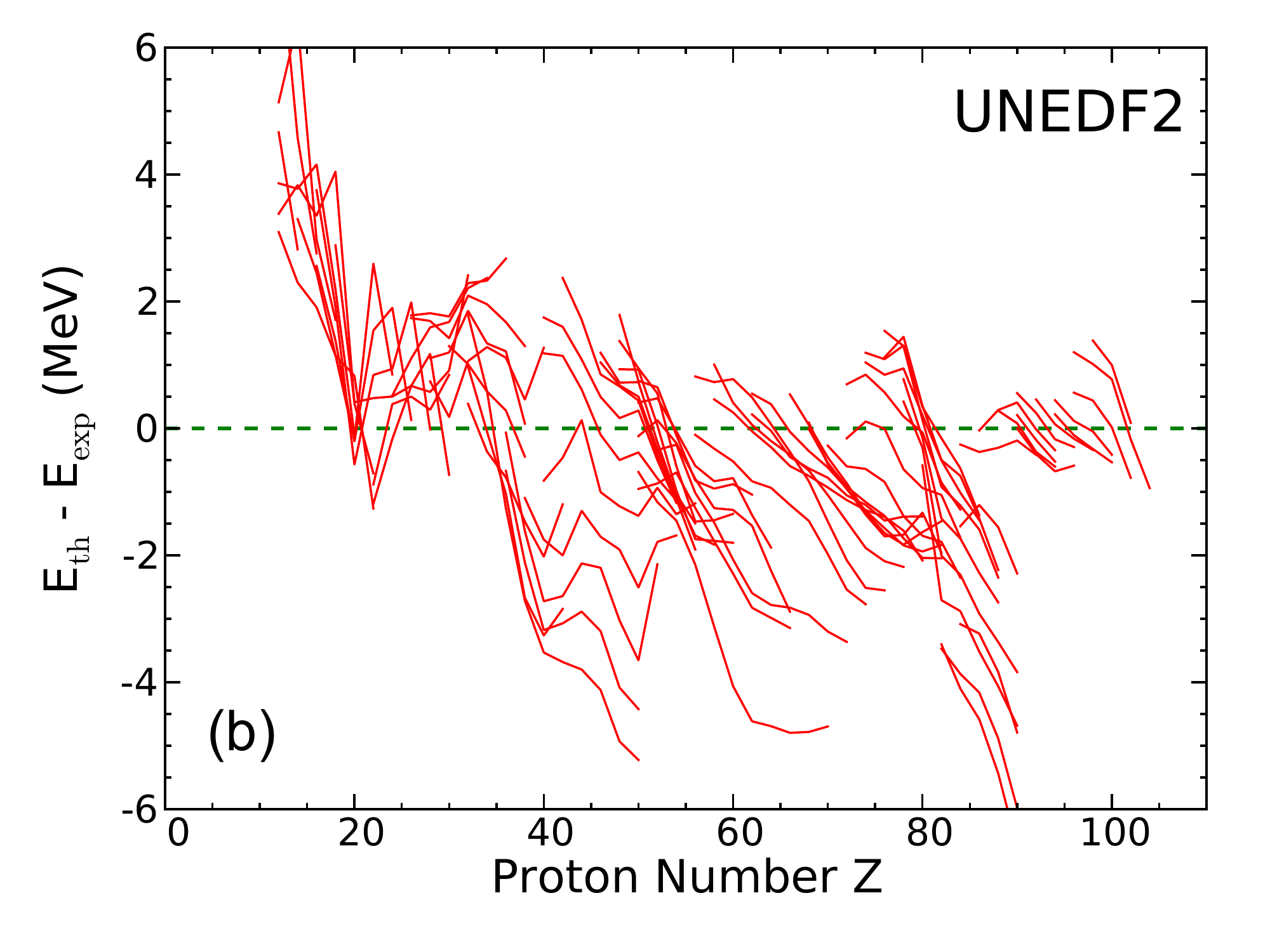}
\caption{(Color online) The residuals of nuclear binding energies of even-even
nuclei calculated with {\UNEDFTWO}. Panel (a) shows isotopic chains, panel
(b) the isotonic chains.}
\label{fig:masstableE}
\end{figure}

\begin{figure}[!htb]
\includegraphics[width=0.9\linewidth]{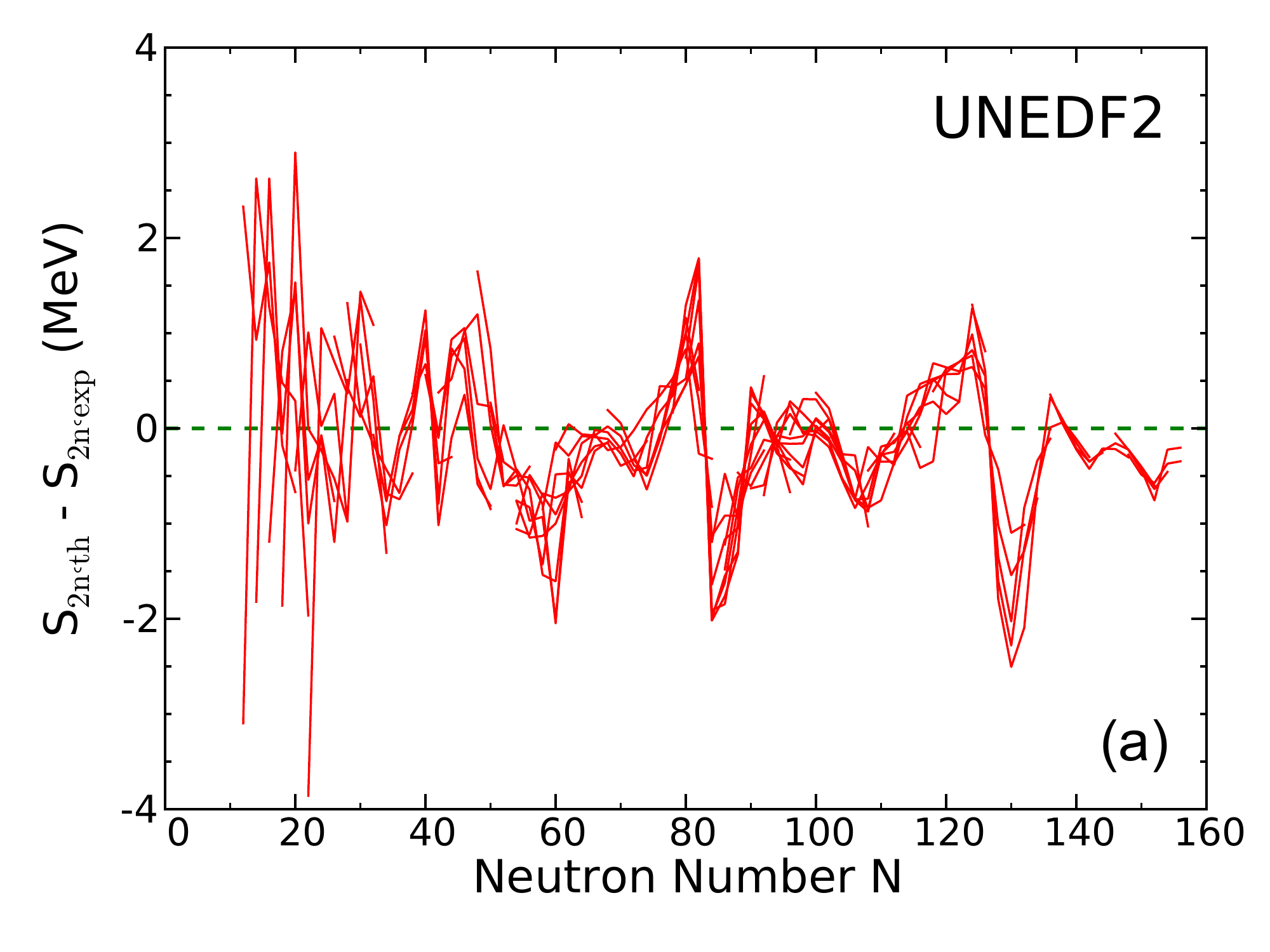} \\
\includegraphics[width=0.9\linewidth]{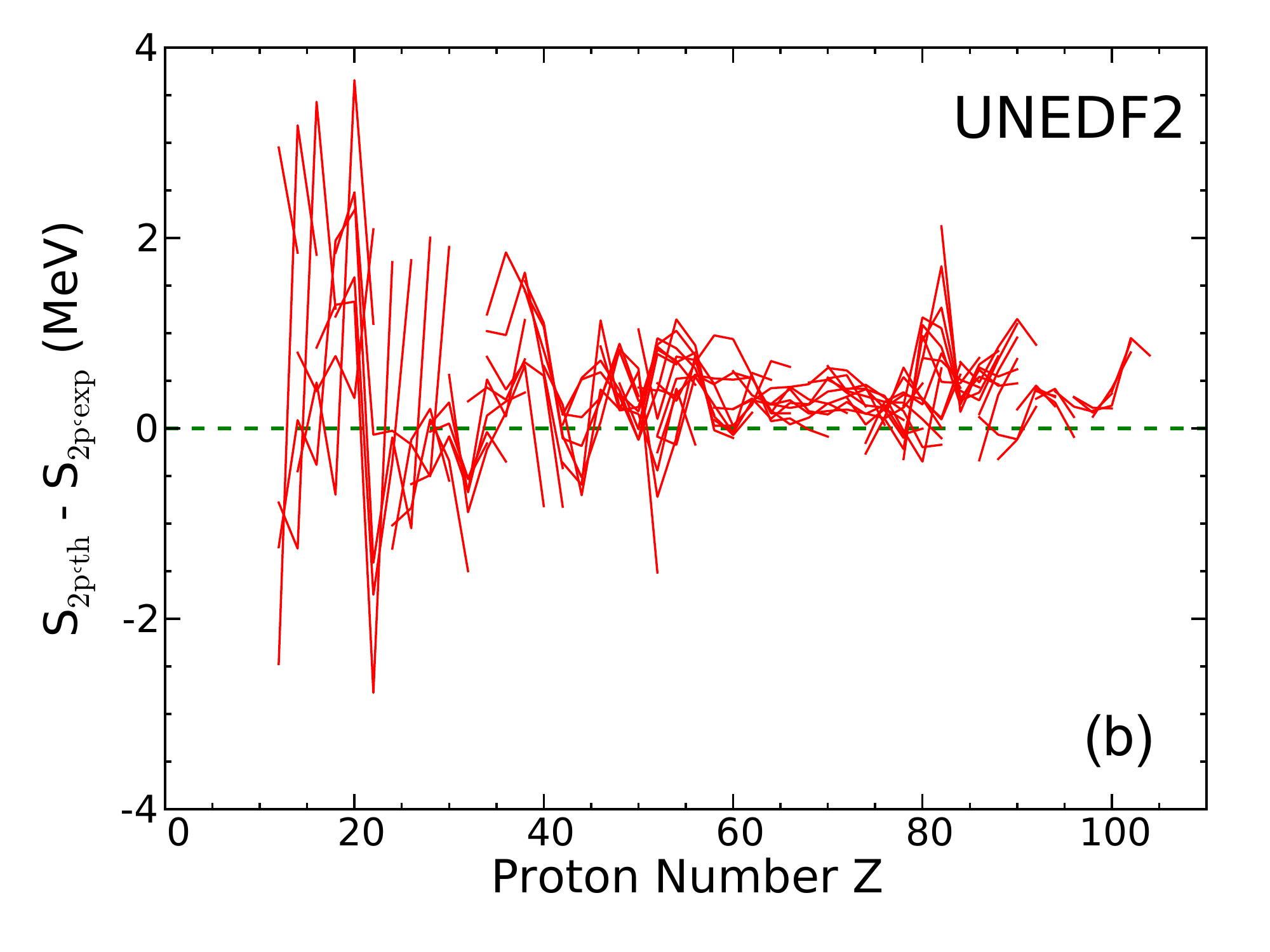}
\caption{(Color online) The residuals of (a) $S_{2n}$ and (b)
$S_{2p}$ obtained in {\UNEDFTWO}
for even-even nuclei.}
\label{fig:separationE}
\end{figure}

\subsection{Global Mass Table}
\label{subsec:masses}

The ability to reproduce nuclear properties globally across the whole nuclear
landscape is one of the key requirements for an universal nuclear EDF. We have
calculated the {\UNEDFTWO} nuclear mass table using the deformed HFB framework
outlined in Ref.~\cite{(Erl12b)}. Figure~\ref{fig:masstableE} shows the
residuals of the nuclear binding energies calculated with {\UNEDFTWO} with
respect to the experimental values for isotopic and isotonic chains of
even-even nuclei. Whereas the residuals for the isotopic chains show the
typical arclike features common to many EDF calculations, these are hardly
present in the isotonic chain residuals. It is difficult to explain this
result, which may point to beyond mean-field effects not included in our
functional and the related bias of the optimization \cite{(Ben06)}.

Figure~\ref{fig:separationE} shows the residuals obtained in {\UNEDFTWO} for
two-neutron and two-proton separation energies. When compared with the
prediction of {\UNEDFONE} \cite{(Kor12)}, the slightly worse RMSD reported in
Table \ref{table:masstable} primarily comes from larger deviations at the ends
of each isotopic chain. As far as $S_{2p}$ values are concerned, {\UNEDFONE}
yields values that are systematically too high. This trend is much less
pronounced with {\UNEDFTWO}.

Table \ref{table:masstable} lists the RMSDs for binding energies, two-particle
separation energies, pairing gaps, and proton radii of even-even nuclei.
Compared with {\UNEDFONE}, {\UNEDFTWO} is slightly less predictive for binding
energies, $S_{2n}$ values, and proton radii, but offers better reproduction
of two-proton separation energies and neutron pairing gaps. The differences
are, however, small.

\begin{table}[!htb]
\caption{\label{table:masstable} RMSDs from  experiment for various observables
calculated with {\UNEDFZERO}, {\UNEDFONE}, and {\UNEDFTWO}. The last column
gives the number of data points used to compute the RMSD.}
\begin{ruledtabular}
\begin{tabular}{lcccr}
Observable &                            {\UNEDFZERO} & {\UNEDFONE} & {\UNEDFTWO} & No. \\
\hline
$E$                                    & 1.428 & 1.912 & 1.950 & 555 \\
$E$ $(A<80)$                           & 2.092 & 2.566 & 2.475 & 113 \\
$E$ $(A\ge 80)$                        & 1.200 & 1.705 & 1.792 & 442 \\[4pt]
$S_{2\rm n}$                           & 0.758 & 0.752 & 0.843 & 500 \\
$S_{2\rm n}$ $(A<80)$                  & 1.447 & 1.161 & 1.243 & 99 \\
$S_{2\rm n}$ $(A\ge 80)$               & 0.446 & 0.609 & 0.711 & 401 \\[4pt]
$S_{2\rm p}$                           & 0.862 & 0.791 & 0.778 & 477 \\
$S_{2\rm p}$ $(A<80)$                  & 1.496 & 1.264 & 1.309 & 96 \\
$S_{2\rm p}$ $(A\ge 80)$               & 0.605 & 0.618 & 0.572 & 381 \\[4pt]
$\tilde{\Delta}_{\rm n}^{(3)}$             & 0.355 & 0.358 & 0.285 & 442 \\
$\tilde{\Delta}_{\rm n}^{(3)}$ $(A<80)$    & 0.401 & 0.388 & 0.327 & 89 \\
$\tilde{\Delta}_{\rm n}^{(3)}$ $(A\ge 80)$ & 0.342 & 0.350 & 0.273 & 353 \\[4pt]
$\tilde{\Delta}_{\rm p}^{(3)}$             & 0.258 & 0.261 & 0.276 & 395 \\
$\tilde{\Delta}_{\rm p}^{(3)}$ $(A<80)$    & 0.346 & 0.304 & 0.472 & 83 \\
$\tilde{\Delta}_{\rm p}^{(3)}$ $(A\ge 80)$ & 0.229 & 0.248 & 0.194 & 312 \\[4pt]
$R_{\rm p}$                            & 0.017 & 0.017 & 0.018 & 49 \\
$R_{\rm p}$ $(A<80)$                   & 0.022 & 0.019 & 0.020 & 16 \\
$R_{\rm p}$ $(A\ge 80)$                & 0.013 & 0.015 & 0.017 & 33
\end{tabular}
\end{ruledtabular}
\end{table}


\subsection{Fission Barriers and Deformation Properties}
\label{subsec:fission}

One of the major differences between the original version of the {\UNEDF}
optimization protocol, used to determine the {\UNEDFZERO} parameterization, and
its successive incarnations used to produce {\UNEDFONE} and {\UNEDFTWO}, is the
inclusion of data on fission isomer excitation energies. This was motivated by
the realization that surface properties of the energy density play a critical
role in the EDF's ability to predict fission properties such as barriers and,
consequently, spontaneous fission half-lives \cite{(Bar82),(Ton85),(Ber89)}. It
was later shown that adding  data corresponding to large nuclear deformations
provides an effective constraint on the surface terms \cite{(Nik10)}.

\begin{figure*}[!htb]
\includegraphics[width=0.75\linewidth]{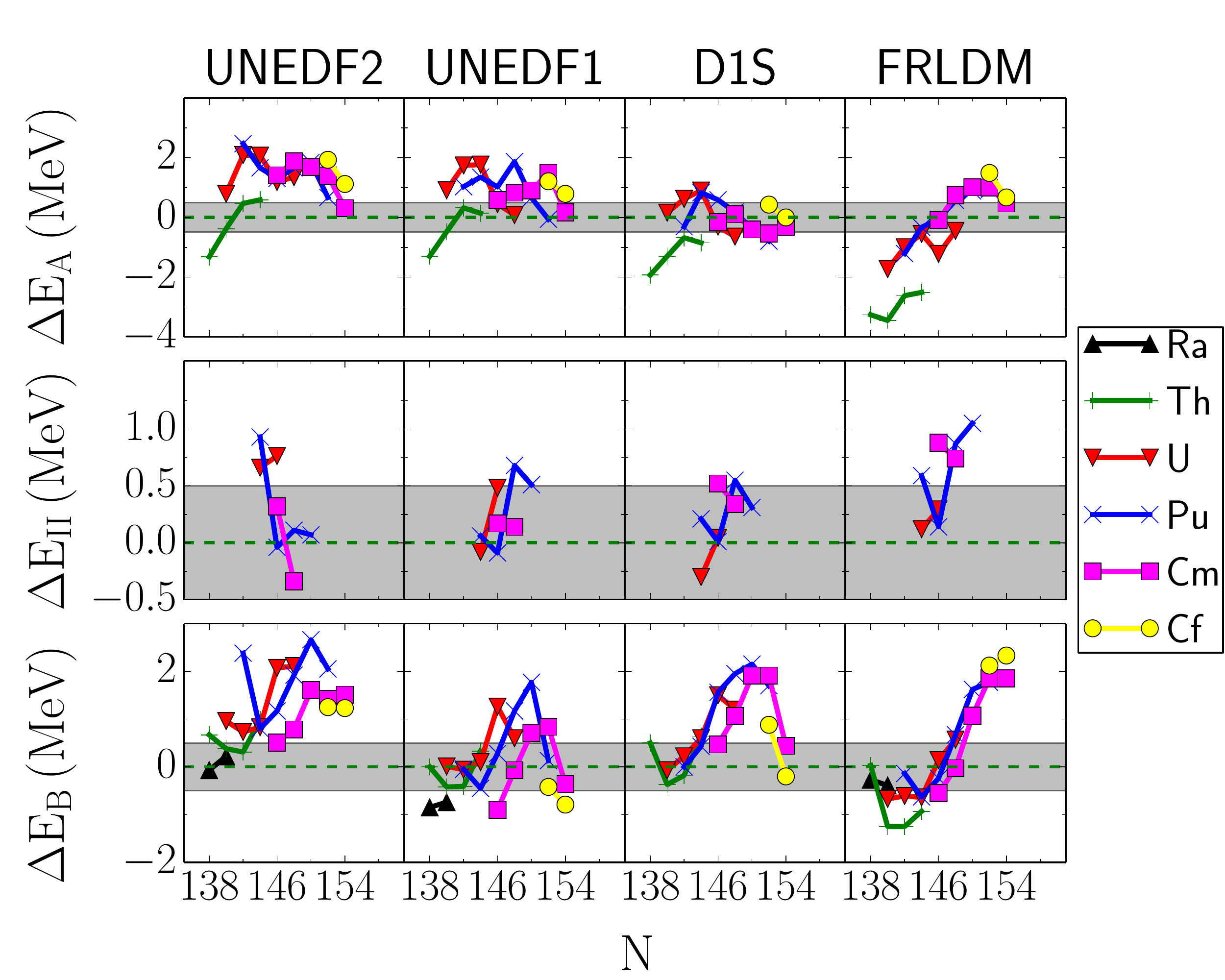}
\caption{(Color online) The residuals of the inner fission barriers,
$\Delta E_{A}$, panels (a)-(d); fission isomer excitation energies, $\Delta E_{II}$,
panels (e)-(h); and outer fission barriers, $\Delta E_{B}$, panels (i)-(l), for various
actinide nuclei. Residuals are defined as the difference between the computed values with
{\UNEDFTWO}, {\UNEDFONE}, D1S, and FRLDM models and the empirical values \cite{(Smi93),(Sin02)}. The shaded area represents an average experimental uncertainty
for each quantity.}
\label{fig:fission}
\end{figure*}

In Fig.~\ref{fig:fission}, we present the residuals for the inner fission
barrier heights,  fission isomer excitation energies, and  outer fission
barrier heights in the actinide region calculated with {\UNEDFONE},
{\UNEDFTWO}, the Gogny D1S model~\cite{(Ber89)}, and the Finite-Range Liquid
Droplet Model (FRLDM) \cite{(Mol09)}. Although  excitation energies of
fission isomers are observables, fission barriers are not. Furthermore, the
uncertainty on the empirical barrier heights ranges from $0.3$\,MeV \cite{(Smi93)} to $1$\,MeV,
while the uncertainty for fission isomer energies ranges from $0.5$\,keV for $^{238}$U
to $0.5$\,MeV for $^{240}$Pu (due to two different values reported in the literature) \cite{(Sin02)}.
To keep the figure legible while conveying information on
experimental uncertainties, the shaded area shows the average empirical
error over the isotopes considered. All
calculations were performed with the DFT solver {\HFODD} of
Ref.~\cite{(Sch11)}. Details of the numerical implementation are discussed in
Refs.~\cite{(McD13),(Sch13)}.

As seen in Fig.~\ref{fig:fission}, the deformation properties of the
{\UNEDFTWO} functional are slightly degraded as compared to  {\UNEDFONE},
especially for the outer barrier. The overall trend is that both barrier
heights tend to be overestimated. This is quantified in Table
\ref{table:fission}, which lists the calculated RMSDs for the calculated first
and second barrier heights, and   fission isomer bandheads. The deviation from
empirical values  has increased by nearly 50\% for the first barrier, and has
doubled for the second barrier. The overall quality of {\UNEDFTWO} is now
comparable to the SkM* parameterization \cite{(Bar82)}.

\begin{table}[!htb]
\caption{\label{table:fission} The RMSDs for the inner barrier height
$E_{\rm A}$, fission isomer bandhead $E_{\rm II}$, and inner barrier height
$E_{\rm B}$ calculated with {\UNEDFONE}, {\UNEDFTWO}, SkM*~\cite{(Bar82)} and
FRLDM \cite{(Mol09)} for the selected even-even actinides (in MeV).}
\begin{ruledtabular}
\begin{tabular}{lrrrrr}
& {\UNEDFTWO} & {\UNEDFONE} & FRLDM & SkM* & D1S \\
\hline
$E_{\rm A}$  & 1.470 & 1.030 & 1.520 & 1.610 & 0.709 \\
$E_{\rm II}$ & 0.515 & 0.357 & 0.675 & 0.351 & 0.339 \\
$E_{\rm B}$  & 1.390 & 0.690 & 1.130 & 1.390 & 1.140
\end{tabular}
\end{ruledtabular}
\end{table}

As discussed in Ref.~\cite{(Nik10)}, the surface and surface-symmetry
coefficients of the leptodermous expansion of the nuclear energy determine
average deformation properties of EDFs at large neutron-proton asymmetries.
Table \ref{table:nucmat} lists the coefficients of the liquid drop expansion
extracted for the three {\UNEDF} functionals, determined according to the
methodology of Ref.~\cite{(Rei06)}. We remark that the surface and curvature
coefficients of both {\UNEDFONE} and {\UNEDFTWO} are very similar. However, the
surface-symmetry coefficient is significantly larger for the {\UNEDFTWO}
parametrization, and takes a value that is comparable to that of {\UNEDFZERO}
and SkM*. This result explains why fission barriers (especially the outer
barrier) are overestimated and similar to what can be obtained with SkM*.

\begin{table}[!htb]
\caption{\label{table:nucmat} Liquid drop coefficients (in MeV) of {\UNEDF}
and SkM*.}
\begin{ruledtabular}
\begin{tabular}{cccccc}
Functional   & $a_{\rm vol}$ & $a_{\rm sym}$  & $a_{\rm surf}$ & $a_{\rm curv}$& $a_{\rm ssym}$ \\
\hline
{\UNEDFZERO} & $-$16.056 & 30.543    & 18.7 & 7.1 & $-$44  \\
{\UNEDFONE}  & $-$15.800 & 29.987    & 16.7 & 8.8 & $-$29  \\
{\UNEDFTWO}  & $-$15.800 & 29.131    & 16.8 & 8.7 & $-$42  \\
SkM*         & $-$15.752 & 30.040    & 17.6 & 9.0 & $-$52
\end{tabular}
\end{ruledtabular}
\end{table}

It also suggests a complex interplay between shell effects and bulk properties
that the EDF optimization has difficulties in keeping under control. As is
well known, the spherical shell
structure plays a major role in driving deformation properties \cite{(Nil95)}. Looking back at
Fig.~\ref{fig:spe208n}, we see that the positions of the neutron 1$j_{15/2}$
and proton 1$i_{13/2}$ shells in {\UNEDFTWO} are depleted as compared to
experiment and {\UNEDFONE}. These high-$j$ orbitals are especially sensitive to
the surface terms of the functional and play an essential role in determining
deformation properties of actinides.


\subsection{Neutron Droplets}
\label{subsec:droplets}

Trapped neutron droplets  constitute a useful theoretical laboratory to test
various many-body methods and effective interactions in inhomogeneous neutron
matter. In particular, they  probe the isovector channels of interactions or
functionals, the role of which increases with neutron excess. The physics of
neutron-rich nuclei is particularly relevant in the context of the inner crust
of neutron stars \cite{(Rav83)}, the $r$-process of nucleo-synthesis
\cite{(Rav83)}, and the determination of the limits of nuclear stability
\cite{(Erl12),(Erl13)}.

Since pure neutron matter is not self-bound, the neutron droplet must be
confined by an external potential in order to produce bound states
\cite{(Pie03)}. Recently, trapped neutron droplets  have been used to test various
ab initio approaches against DFT calculations with phenomenological functionals
\cite{(Gan11)}. In particular, in Ref.~\cite{(Bog11)}, neutron droplets were
used to test density matrix expansion techniques, which aim at building EDFs
from the realistic interactions used in ab initio methods.

\begin{figure}[!htb]
\includegraphics[width=0.9\linewidth]{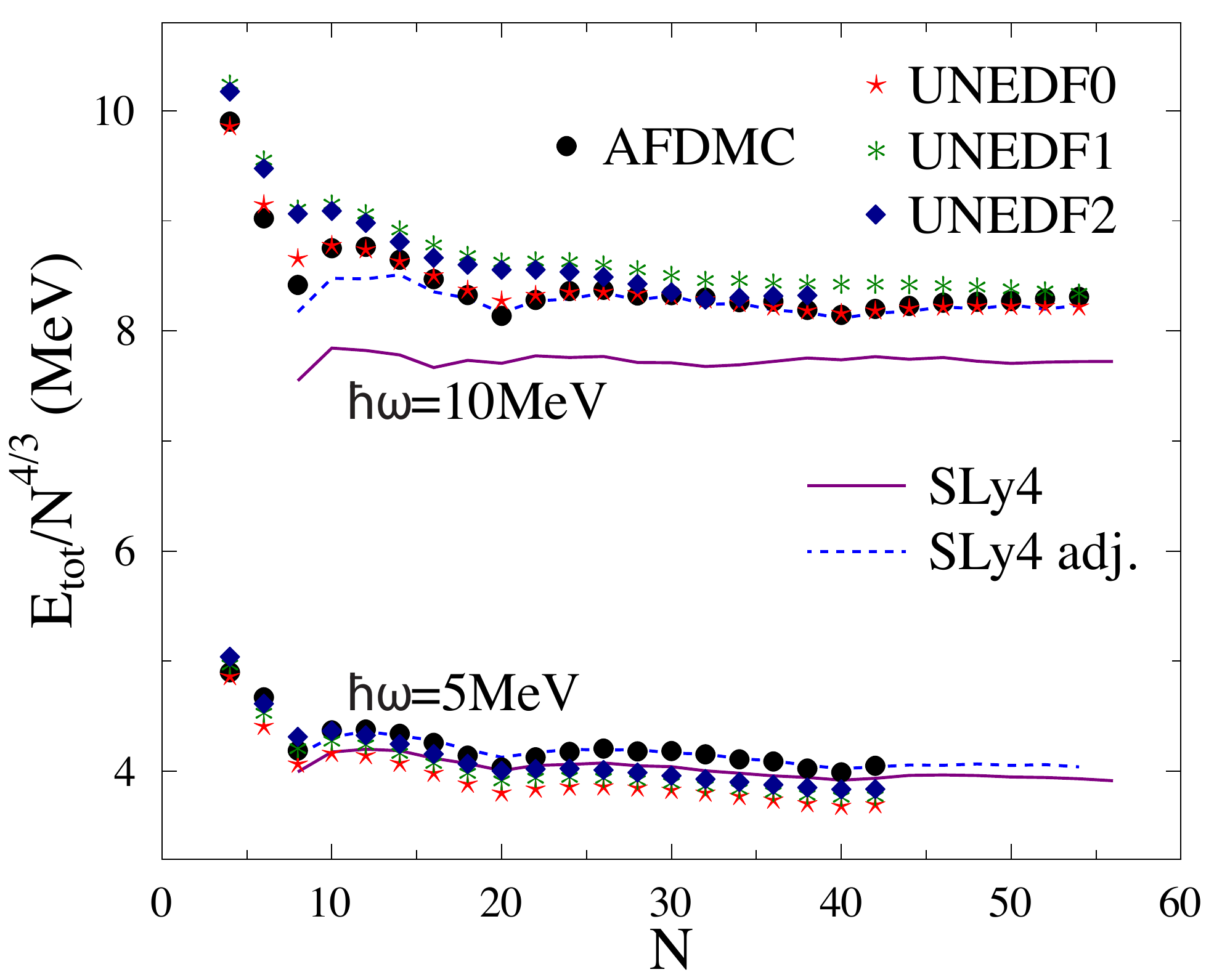}
\caption{(Color online) Neutron droplet energies predicted with {\UNEDFZERO},
{\UNEDFONE}, and {\UNEDFTWO} compared to the ab-initio AFDMC results and DFT
calculations with SLy4 and adjusted SLy4 EDFs of Ref.~\cite{(Gan11)}.}
\label{fig:ndroplets}
\end{figure}

In Fig.~\ref{fig:ndroplets}, the binding energy per neutron of neutron droplets
calculated with {\UNEDFZERO}, {\UNEDFONE}, and {\UNEDFTWO} are compared with
the ab initio results obtained in Ref.~\cite{(Gan11)} within the Auxiliary
Field Diffusion Monte-Carlo (AFDMC)  method. AFDMC calculations were performed
with the AV8' parameterization of the two-body potential and the Urbana IX
three-body interaction \cite{(Pud97),*(Pud96)}. The figure also shows DFT
calculations with SLy4, as well as a modified SLy4 parameterization that has
been slightly readjusted in the isovector channel to reproduce the AFDMC results.
All neutron droplet systems considered in Fig.~\ref{fig:ndroplets} were
confined by a spherical HO potential, with two choices of the
oscillator frequency, $\hbar\omega = 5$ MeV and $\hbar\omega = 10$ MeV.
As previously seen for
{\UNEDFZERO} and {\UNEDFONE} \cite{(Kor12)}, {\UNEDFTWO} results are close to
the ab initio calculations, even though the optimization did not include any
information about neutron droplets.

However, we notice that the results for $N>38$ with $\hbar\omega=10$\,MeV are
not available for {\UNEDFTWO}. This situation is the direct consequence of the
neutron matter instabilities discussed in Sec.~\ref{subsec:linearresponse}.
For $N>38$ droplets, the central neutron density exceeds the critical density
shown in Fig.~\ref{fig:pnm}; as a result, the HFB  calculation fails to
converge. For $\hbar\omega=5$\,MeV, the central neutron density is low enough
for higher particle numbers, so that the instabilities do not appear.


\section{Conclusions}
\label{sec:conclusions}

In this study, we have introduced the {\UNEDFTWO} parameterization of the
Skyrme energy density. Compared with our previous work, there are two main
differences: (i) we released the requirement that the isoscalar and isovector
tensor coupling constants be zero, and (ii) we included experimental data on
s.p.\ level splittings in doubly magic nuclei to better constrain spin-orbit
and tensor coupling constants. In addition to these major changes, we have
slightly extended our dataset to improve the pairing properties of the
functional, especially in heavy nuclei. Following previous {\UNEDF}
optimizations, we have performed a comprehensive sensitivity analysis of our
parameterization in order to obtain standard deviations and correlations among
EDF parameters.

Global nuclear properties computed with {\UNEDFTWO} reflect little or no
improvement with respect to our previous parameterizations. While the linear
response analysis  has shown that {\UNEDFTWO} does not have any finite-size
instabilities in symmetric nuclear matter for densities up to
$1.5\rho_{\rm c}$, some instabilities are encountered in pure neutron matter,
with the consequence that neutron droplet calculations do not converge at large
neutron numbers and large oscillator frequencies. The position of the GDR peak
in $^{208}$Pb is slightly too low in energy, which is attributed to a
persistent lack of constraints on the isovector effective mass. The quality of
the single-particle shell structure near closed shell nuclei is almost as good
as one can get with Skyrme EDFs, but this was almost the case with {\UNEDFZERO}
and {\UNEDFONE}. The RMSD for nuclear binding energies is 1.95 MeV, which is
far from the performance of semi-phenomenological mass models (see, for
example, Ref.~\cite{(Gor13)} for the most recent numbers) and comparable to
{\UNEDFONE}. Deformation properties, which had been significantly improved with
{\UNEDFONE} are degraded markedly for {\UNEDFTWO}, which yields fission
barriers similar to that of the traditional SkM* functional.

On the other hand, as discussed in Sec.~\ref{subsec:UNEDF2}, the interval of
confidence for the parameters is narrower for {\UNEDFTWO} than it was for
{\UNEDFONE}, which itself was more tightly constrained than {\UNEDFZERO}. In
addition, the results of the sensitivity analysis of
Sec.~\ref{subsec:sensitivity} show that there is relatively weak dependence on
individual experimental points. These results point to the fact that the
coupling constants of the {\UNEDFTWO} functional are properly constrained by
the data.

Although one can certainly improve the optimization protocol, for example by
changing the relative weights in the $\chi^{2}$ objective function, we believe
this relative lack of improvement should be viewed as an intrinsic limitation
of the Skyrme energy density, a local energy density that is  up to second
order in derivatives \cite{(Dob96),(Per04)}. Indeed, as shown in
Figs.~\ref{fig:masstableE}-\ref{fig:fission}, the residuals of various
quantities predicted with {\UNEDFTWO} do not have a statistical distribution;
hence, adding more data points or playing with the $\chi^{2}$ is not going to
change the situation as the deviations are mainly affected by systematic
errors, i.e., imperfect modeling. In this context, {\UNEDFTWO} is an all-around
Skyrme EDF that is fairly well constrained by various data, but it also marks
the end of the Skyrme EDF strategy.

At this phase of nuclear DFT developments, it thus seems urgent to go beyond
traditional Skyrme functionals. Two major avenues are being explored: one following the
spirit of DFT, where the primary building block is the energy density
functional that includes all correlation effects, and the other following the
spirit of the self-consistent mean-field theory, where the major ingredient is
an effective pseudopotential and the beyond-mean-field correlations are added
afterwards. The DFT description is especially convenient for tying in the
energy density to a more fundamental theory of nuclear forces based, for
example, on the chiral effective field theory. This can be accomplished by using EDF
built from the density matrix expansion of realistic interactions
\cite{(Sto10),(Bog11),(Geb10),(Car10)}. A complementary route is to explore
functionals with higher order derivatives of the density
\cite{(Car08),(Rai11),(Dob12)}. These EDFs are much richer than the Skyrme or
Gogny functionals; hence, they should be able to capture more physics and
reduce systematic errors.

\bigskip
\begin{acknowledgments}
We are deeply indebted to the late M. Stoitsov, whose contribution to this
work, especially the DFT solver and its interface with the {\algo} algorithm,
was considerable. This work was supported by the U.S. Department of Energy
under Contract Nos. DE-SC0008499, DE-FG02-96ER40963, and DE-FG52-09NA29461
(University of Tennessee), No. DE-AC02-06CH11357 (Argonne National Laboratory),
and DE-AC52-07NA27344 (Lawrence Livermore National Laboratory); by the Academy
of Finland under the Centre of Excellence Programme 2012--2017 (Nuclear and
Accelerator Based Physics Programme at JYFL) and FIDIPRO programme; and by the
European Union's Seventh Framework Programme ENSAR (THEXO) under Grant No.
262010.
Computational resources were provided through an INCITE award ``Computational
Nuclear Structure'' by the National Center for Computational Sciences (NCCS)
and National Institute for Computational Sciences (NICS) at Oak Ridge National
Laboratory, through an award by the Livermore Computing Resource Center at
Lawrence Livermore National Laboratory, and through an award by the Laboratory
Computing Resource Center at Argonne National Laboratory.
\end{acknowledgments}


\bibliographystyle{apsrev4-1}
\bibliography{unedf2}

\end{document}